\documentclass[12pt]{article}
\usepackage[square]{natbib} 
\usepackage{a4}
\usepackage{latexsym} 
\usepackage{amssymb}  
\usepackage{graphicx}
\usepackage{epsfig}

\newcounter{zyxabstract}     
\newcounter{zyxrefers}        

\newcommand{\newabstract}
{\newpage\stepcounter{zyxabstract}\setcounter{equation}{0}
\setcounter{footnote}{0}}

\newcommand{\rlabel}[1]{\label{zyx\arabic{zyxabstract}#1}}
\newcommand{\rref}[1]{\ref{zyx\arabic{zyxabstract}#1}}
\def\si{^1 \hskip -0.03in S _0}
\def\siii{^3 \hskip -0.025in S _1}

\def\nLi{$^7\mathrm{Li}(n, \gamma)^8\mathrm{Li}$}
\def\pBe{$^7\mathrm{Be}(p, \gamma)^8\mathrm{B}$}

{\section*{References}\setcounter{zyxrefers}{0}
\begin{list}
{[\arabic{zyxrefers}]}
{\usecounter{zyxrefers}\setlength{\parindent}{0cm}\setlength{\itemsep}{0cm}}

}
{\end{list}}
\newenvironment{thebibliographynotitle}[1] 
{\section*{References}\setcounter{zyxrefers}{0}
\begin{list}{[\arabic{zyxrefers}]}
{\usecounter{zyxrefers}\setlength{\parindent}{0cm}\setlength{\itemsep}{-1.5mm}}}
{\end{list}}

\renewcommand{\bibitem}[1]{\item\rlabel{y#1}}
\renewcommand{\cite}[1]{[\rref{y#1}]}      

\begin{document}
\begin{titlepage}
\begin{flushright} 
{\small
}
\end{flushright}

\begin{center}
{\LARGE\bf Nuclear Dynamics\\[2.1mm]with Effective Field Theories$^{*}$}
\\[1cm]
Ruhr-Universit\"at Bochum, Germany\\
July 1 -- 3, 2013\\[1cm]
{\bf Evgeny Epelbaum,} {\bf Hermann Krebs}
\\[0.3cm]
Institut f\"ur Theoretische Physik II, Ruhr-Universit\"at Bochum,\\
44780 Bochum, Germany\\[1cm]
{\sl Dedicated to the memory of Professor Walter Gl\"ockle}\\[0.75cm]
{\large ABSTRACT}
\end{center}
These are the proceedings of the international workshop on ``Nuclear
Dynamics with Effective Field Theories''
held at Ruhr-Universit\"at  Bochum, Germany from 
July 1 to 3, 2013. The workshop focused on effective field
theories of low-energy QCD, chiral perturbation theory for nuclear
forces as well as few- and many-body physics.  
Included are a short contribution per talk.

\vfill
\noindent\rule{6cm}{0.3pt}\\
\footnotesize{$^*$ This workshop was supported by the ERC Grant No.~259218
``Nuclear Physics from Quantum Chromodynamics'' (acronym
NuclearEFT). 
This research is part of the EU Research Infrastructure Integrating Activity 
``Study of Strongly Interacting Matter''
(acronym HadronPhysics3).
}
\end{titlepage}

\section{Introduction}

\hskip 0.6 true cm 
The past decade has witnessed a considerable progress towards
understanding nuclear structure and dynamics from first principles. 
In addition to new experimental results,  this is to a
large extent related to exciting theoretical developments in this
field. 
On the one hand, rapidly increasing computational resources coupled
with sophisticated few- and many-body methods allow nowadays for
reliable and accurate nuclear structure calculations for light and
medium-mass nuclei. This opens the possibility to relate the properties of
the nuclear Hamiltonian to observables in a reliable way without
invoking any uncontrollable approximations. On the other hand,
significant progress has been reached towards quantitative
description of nuclear forces and currents in the framework of chiral
effective field theory. Last but not least, first lattice-QCD results for
few-nucleon systems have also been reported.  The main 
goal of this workshop was to bring together experimentalists and
theorists with expertise in these fields to discuss recent
developments and future research directions. 
More specifically, the workshop focused on effective field
theories of low-energy QCD, chiral perturbation theory for nuclear
forces as well as few- and many-body physics.   

This workshop took place in July 2013 at Ruhr-University Bochum,
Germany. It brought together 41 participants whose names, institutions and email addresses
are listed below. 34 of them presented results in talks of 35 minutes in
length. A short description of their contents and a list of the most relevant
references can be found below. We felt that such mini-proceedings
represent a more appropriate framework than full-fledged proceedings. 
Most results are or will soon be published and available on the archive,
so this way we can achieve speedy publication and avoid duplication of
results in the archive. Below we give the program of the workshop, 
followed by the list of participants and, finally, by the abstracts of
all the talks. The slides of the talks are available at the 
workshop website  

\noindent
{\small \tt  http://www.tp2.ruhr-uni-bochum.de/$\sim$ndeft13/}

We extend our sincere gratitude to the European Research Council (ERC)
for the financial support and to Martina Hacke and Peter Druck for the
precious help with the administration and organization. We are also
grateful to  many students of the Institute of Theoretical Physics II
and especially to Arseniy Filin, Dmitrij Siemens and  Markus
Th\"urmann 
for their help during the
workshop and to the team of the RUB Veranstaltungszentrum for the
excellent service. Most important, we thank all the participants for making this an exciting
and lively meeting happen.

This workshop was dedicated to Professor Walter Gl\"ockle, an esteemed
colleague, advisor and friend of many in our field, who passed away on August 1,
2012. 

\vspace{0.5cm}

\hfill Evgeny Epelbaum and  Hermann Krebs

\newpage

\section{Program}
\begin{tabbing}
xx:xx \= A very very very long name \= \kill
{\bf \large Monday, July 1st, 2013}\\[6pt]
09:20 \>    Evgeny Epelbaum    \>  Welcome  remarks \\
\> (Bochum) \\ [10pt]
{\it  Early Morning Session}\\
Chair: Charlotte Elster    \>  \> {\bf  Few-Body I} \\
  09:35 \>    Wayne Polyzou      \>  
Walter and the relativistic few-body problem \\
         \> (Iowa City) \>      \\
10:10 \>      Henryk Wita{\l}a (Krakow) \> 
  3N reactions with N$^3$LO chiral force \\
10:45 \>\>{\em  Coffee}\\[10pt]
{\it   Late Morning Session}\\
Chair: Charlotte Elster    \>  \> {\bf  Few-Body II} \\
11:15 \>
        Alejandro Kievsky (Pisa)\>   Scattering states from bound
        state like wave functions: \\
          \> \> The Coulomb case      \\
11:50 \>
        Hideki Sakai  (Tokyo)  \>
       Story of how we started $dp$ scattering experiments at RIKEN \\ 
12:25\>\>{\em End of Session}\\
12:25\>\>{\em Lunch}\\[10pt]
{\em Early Afternoon Session}\\
        Chair: Charlotte Elster \>\>{\bf   Few-Body III}\\
14:00\>
        Hans P\"atz gen.~Schieck     \>  Spin physics
        and "polarized fusion"   \\
\> (K\"oln) \> \\
14:35\>
        Werner Tornow \> Are new experiments really
        needed to advance few-body\\
   \> (Duke, USA)   \>  physics?  \\
15:10\>\>{\em Coffee}\\[10pt]
{\em Afternoon Session}\\
        Chair: Dean Lee  \>\>{\bf QCD and chiral dynamics I  }\\
15:30\>
        Silas Beane (Bonn)  \>   Nuclear dynamics from lattice QCD \\
16:05\>
        Veronique Bernard (Orsay)   \> SU(3) chiral dynamics revisited
        \\
16:40\>\>{\em Coffee}\\[10pt]
{\em Late Afternoon Session}\\
        Chair: Dean Lee  \>\>{\bf QCD and chiral dynamics II  }\\
17:00\>
        Christian Weiss (JLab)  \>    	Space-time picture of chiral dynamics with nucleons  \\
17:35\>
        Vadim Baru (Bochum)   \>  Pion production in NN collisions    \\
18:10\>
        Jambul Gegelia (Bochum)   \>  Issues of renormalization in EFT
        for NN system    \\
18:45\>\>        
{\em End of Session}\\
[36pt]
{\bf \large Tuesday, July 2nd, 2013}\\[6pt]
{\em Early Morning Session}\\
        Chair: Evgeny Epelbaum \> \> {\bf Nuclear Lattice EFT}\\
09:00\>
        Dean Lee (Raleigh)  \>  Overview and latest news from nuclear
        lattice effective \\
   \> \>     field theory \\
09:35\>
        Ulf-G.~Mei{\ss}ner  \>  Life on Earth -- an accident? \\
   \>  (Bonn/J\"ulich)\>   \\
10:10\>
        Gautam Rupak \> Nuclear structure and reactions in lattice effective \\
   \>  (Mississippi State) \>     field theory\\
10:45\>\>        
{\em Coffee}\\[10pt]
{\em Late Morning Session}\\
        Chair: Evgeny Epelbaum  \>\>{\bf  Many-Body I}\\
11:15\>
        James Vary (Ames, Iowa) \>  Evolving perspectives on the
        origins of nuclear structure  \\
11:50\>
        Robert Roth (Darmstadt) \>  From chiral EFT interactions to nuclear structure and back  \\
12:25\>\>{\em End of Session}\\
12:25\>\>{\em Lunch}\\[10pt]
{\em Early Afternoon Session}\\
        Chair: James Vary  \> \>  {\bf Many-Body II}\\
14:00\>
        Carolina Romero \>  Ab initio many-body calculations
        of light ion reactions   \\
\> (TRIUMF) \> \\ 
14:35\>
        Kai Hebeler (Darmstadt) \>  Neutron rich matter from chiral EFT interactions  \\
15:10\>\>        
{\em Coffee}\\[10pt]
{\em Afternoon Session}\\
        Chair: James Vary    \> \>{\bf Halo-Nuclei}\\
15:30\>
        Hans-Werner Hammer \>     Universal properties of halo nuclei \\
\> (Bonn) \> \\ 
16:05\>
        Daniel Phillips  \>  Photon interactions with
        halo nuclei  \\
\> (Athens, USA)  \> \\
16:40\>\>        
{\em Coffee}\\[10pt]
{\em Late Afternoon Session}\\
        Chair: James Vary    \> \>{\bf Few-Body Electroweak I}\\
17:00\>
        Jacek Golak (Krakow)   \> Selected weak interaction processes
        on the deuteron and $^3$He  \\
17:35\>
        Harald Grie{\ss}hammer\> High-accuracy analysis of Compton scattering in chiral EFT:   \\
\>  (Washington DC)   \> Status and future \\
18:10\> \>{\em End of Session}\\
19:00\> \>{\em Workshop Dinner}\\[48pt]
{\bf \large Wednesday, July 3rd, 2013}\\[6pt]
{\em Early Morning Session}\\
        Chair: Hermann Krebs   \> \>{\bf  Three-Nucleon Force (Theory)}\\
09:00\>
        Ashot Gasparyan    \>    Chiral expansion of the
        three-nucleon force \\ 
\> (Bochum)  \> \\ 
09:35\>
        Carlos Schat \>  Three-nucleon forces in
        the $1/N_c$ expansion  \\
\> (Buenos Aires)  \> \\
10:10\>
        Luca Girlanda (Lecce)  \>  Constraining the three-nucleon
        contact interaction from \\
\> \> nucleon-deuteron elastic scattering
        \\ 
10:45\>\>{\em Coffee}\\[10pt]
{\em Late Morning Session}\\
        Chair: Hermann Krebs  \>  \>{\bf  Three-Nucleon Force (Experiment)}\\
11:15\>
        Kimiko Sekiguchi (Sendai)   \>   Exploring three nucleon
        forces in few-nucleon scattering \\
11:50\>
        Nasser Kalantar (KVI) \>  Study of nuclear forces at
        intermediate energies  \\
12:25\> \>{\em End of Session}\\
12:25\> \>{\em Lunch}\\[10pt]
{\em Early Afternoon Session}\\
        Chair: Henryk Wita{\l}a   \> \> {\bf Hyper-nuclei}\\
14:00\>
        Andreas Nogga (FZJ)  \>   	Light hypernuclei based on
        chiral interactions at \\
\> \> next-to-leading order  \\
14:35\>
        Kazuya Miyagawa    \>    The reaction $K^- \, d \, \to
        \, \pi \Sigma n$ in the $\Lambda (1405)$ resonance region  \\
\> (Okayama)  \> \\
15:10\>\>{\em Coffee}\\[10pt]
{\em Afternoon Session}\\
        Chair: Henryk Wita{\l}a \>  \>{\bf  Few-Body Electroweak II}\\
15:30\>
        Roman Skibinski (Krakow)    \>    The chiral electromagnetic
        currents applied to the deuteron \\
 \>  \> and $^3$He disintegrations
        \\ 
16:05\>
        Fred Myhrer   \>    Muon capture and the connection to the three-nucleon force 
        \\ 
\>  (Columbia, USA)  \> \\
16:40\> \>        {\em Coffee}\\ [10pt]
{\em Late Afternoon Session}\\
        Chair: Henryk Wita{\l}a \> \> {\bf  Few-Body IV}\\
17:00\>
        Charlotte Elster   \>  Towards $(d,p)$ reactions
        with heavy nuclei in a Faddeev \\ 
\> (Athens, USA)   \>  description  \\
17:35\>
        Stanis{\l}aw Kistryn \>   Experimental studies of
        deuteron-proton breakup at medium \\
\>  (Krakow)   \> energies  \\
18:10\>
        Hiroyuki Kamada (Kyushu)   \>  $Nd$ scattering calculation with low-momentum potential \\ 
18:45 \>  Hermann Krebs (Bochum)  \> Concluding remarks\\
19:00\> \>{\em End of Workshop}
\end{tabbing}

\section{Participants and their email addresses}

\begin{tabbing}
A very long namexxxxxxx\=a very long institutexxxxxxxxx\=email\kill
V. Baru \> Ruhr-Univ. Bochum \>  vadimb@tp2.rub.de \\
S.\,R. Beane \> Univ. Bonn/New Hampshire \> silas@physics.unh.edu \\
V. Bernard \> IN2P3, CNRS \> bernard@ipno.in2p3.fr \\
S. Elhatisari \> NC State Univ. \> selhati@ncsu.edu \\
Ch. Elster \> Ohio Univ. \> elster@ohiou.edu \\
E. Epelbaum \> Ruhr-Univ. Bochum \>  evgeny.epelbaum@rub.de \\
A. Gasparyan \>  Ruhr-Univ. Bochum   \> ashotg@tp2.rub.de \\
J. Gegelia \> Ruhr-Univ. Bochum   \> jambul.gegelia@tp2.rub.de \\
L. Girlanda \> INFN Lecce \> girlanda@le.infn.it \\
J. Golak \> Jagiellonian Univ.  \> ufgolak@cyf-kr.edu.pl \\
H.\,W. Grie{\ss}hammer \> GWU \> hgrie@gwu.edu \\
H.-W. Hammer \> Univ. Bonn \> hammer@hiskp.uni-bonn.de \\
K. Hebeler \> TU Darmstadt \> kai.hebeler@physik.tu-darmstadt.de \\
N. Kalantar-Nayestanaki \> KVI \> nasser@kvi.nl\\
H. Kamada \> Kyushu Inst. Tech. \> kamada@mns.kyutech.ac.jp \\
A. Kievsky \> INFN Pisa \> alejandro.kievsky@pi.infn.it \\
J. Kirscher \>   EuBW Karlsruhe  \>   kirscher@gwmail.gwu.edu \\
S. Kistryn \>  Jagiellonian Univ.  \> stanislaw.kistryn@uj.edu.pl \\
H. Krebs \>  Ruhr-Univ. Bochum \> hermann.krebs@rub.de \\
D. Lee \> NC State Univ. \> dean\_lee@ncsu.edu \\
U.-G. Mei{\ss}ner \> Univ.\ Bonn \& FZ J\"ulich \>
meissner@hiskp.uni-bonn.de \\
D. Minossi \> Ruhr-Univ. Bochum/FZJ \>  d.minossi@fz-juelich.de \\
K. Miyagawa \> Okayama Univ. \> miyagawa@dap.ous.ac.jp \\
F. Myhrer \> SC State Univ. \> myhrer@caprine.physics.sc.edu \\
A. Nogga \> FZ J\"ulich \> a.nogga@fz-juelich.de \\
H. P\"atz gen. Schieck \> K\"oln Univ. \> h.schieck@t-online.de \\
M. Pine \> NC State Univ. \> mjmantoo@ncsu.edu \\
D. Phillips \> Ohio Univ. \> phillips@phy.ohiou.edu \\
M. Polyakov \> Ruhr-Univ. Bochum \>  maxim.polyakov@tp2.rub.de \\
W. Polyzou \> Iowa Univ.  \> polyzou@gelfand.physics.uiowa.edu \\ 
C. Romero-Redondo \> TRIUMF \> cromeroredondo@triumf.ca \\
R. Roth \> TU Darmstadt \> robert.roth@physik.tu-darmstadt.de \\
G. Rupak \> Mississippi State Univ. \> grupak@gmail.com \\
H. Sakai \> RIKEN \> hsakai@ribf.riken.jp \\
C. Schat \> Univ. Buenos Aires  \> carlos.schat@gmail.com \\
K. Sekiguchi \> Tohoku Univ. \> kimiko@lambda.phys.tohoku.ac.jp \\
R. Skibinski \>  Jagiellonian Univ.  \> roman.skibinski@uj.edu.pl \\
W. Tornow \> Duke Univ/TUNL \> tornow@tunl.duke.edu \\
J. Vary \> Iowa State Univ. \> jvary@iastate.edu \\
C. Weiss \> JLab \> weiss@jlab.org \\
H. Wita{\l}a \>  Jagiellonian Univ.  \> ufwitala@cyf-kr.edu.pl 
\end{tabbing}

\newabstract 
\begin{center}
{\large\bf Walter and the relativistic few-body problem}\\[0.5cm]
{\bf W. N. Polyzou}  \\[0.3cm]
Department of Physics and Astronomy\\
The University of Iowa\\
Iowa City IA USA\\[0.3cm]
\end{center}

One of Walter Gl\"ockle's many interests was to apply few-body methods
at relativistic energy scales.  At these scales it is possible to
probe the short distance properties of the nucleon-nucleon and three-nucleon
interactions.  These are scales where sub-nucleon degrees of freedom
may begin to be relevant.  Relativistic methods are also needed for a
consistent treatment of particle production and to model reactions
involving hypernuclei.  Also, because cluster properties generate 
many-body forces, it is important to understand how these forces interact
with conventional many-body forces.  Finally, QCD is strongly coupled
at these scales and is difficult to control mathematically.

Walter, along with his students and collaborators made a significant
amount of scientific progress in the study of relativistic few-nucleon
physics.  His contributions include relativistic models of interacting
nucleons with meson degrees of freedom eliminated\cite{r1},
construction of realistic relativistic interactions \cite{r6},
calculations of relativistic effects on the three-nucleon binding
energy \cite{r3}, \cite{r8}, relativistic effects in proton-deuteron
scattering and the interplay of relativistic effects with three-nucleon
forces \cite{r8}, \cite{r9}, \cite{r10}, the role of relativity in 
Y-scaling \cite{r11}, GeV-scale three-nucleon scattering calculations
without partial waves \cite{r14} and relativistic spin effects 
in low-energy calculations of $A_y$\cite{r16}.

\setlength{\bibsep}{0.0em}
\begin{thebibliographynotitle}{99}
\bibitem{r1} Walter Gl\"ockle and L. M\"uller,
Phys. Rev. {\bf C}23 (1981) 1183-1195.
\bibitem{r6} H. Kamada and W. Gl\"ockle,
Phys. Lett. {\bf B}655 (2007) 119-125.
\bibitem{r3} W. Gl\"ockle, T. S. H. Lee, F. Coester,
Phys. Rev. {\bf C}33 (1986) 709-716.
\bibitem{r8} H. Kamada, W. Gl\"ockle, H. Wita{\l}a, J. Golak, R. Skib\'inski, 
W. Polyzou, Ch. Elster, Mod. Phys. Lett. {\bf A}24 (2009) 804-809.
\bibitem{r9} H. Wita{\l}a, J. Golak, W. Gl\"ockle, H. Kamada, 
Phys. Rev. {\bf C}71  (2005) 054001.
\bibitem{r10} H. Wita{\l}a, J. Golak, R. Skib\'inski, W. Gl\"ockle, H. Kamada, W.N. Polyzou,
Phys. Rev. {\bf C}83 (2011) 044001.
\bibitem{r11} W. N. Polyzou and Walter Gl\"ockle,
Phys. Rev. {\bf C}53 (1996) 3111-3130.
\bibitem{r14} T. Lin, Ch. Elster, W. N. Polyzou, H. Wita{\l}a, W. Gl\"ockle,
Phys. Rev. C78 (2008) 024002.
\bibitem{r16} H. Wita{\l}a, J. Golak, R. Skib\'inski, W. Gl\"ockle, 
W. N. Polyzou, H. Kamada,
Phys. Rev. C77 (2008) 034004.
\end{thebibliographynotitle}

\newabstract 
\begin{center}
{\large\bf 3N reactions with N$^3$LO chiral force}\\[0.5cm]
{\bf Henryk Wita{\l}a}  \\[0.3cm]
M. Smoluchowski Institute of Physics, Jagiellonian
University,  PL-30059 Krak\'ow, Poland \\[0.3cm]
\end{center}

Comparison of theoretical predictions with
data for elastic nucleon-deuteron (Nd) scattering and nucleon
induced deuteron breakup clearly shows the importance of the three-nucleon
force (3NF).  
Inclusion of semi-phenomenological 3NF models  into calculations
 in many cases  improves the data description. However, some serious 
discrepancies remain even when 3NF is included.

At low energies the prominent examples were found for the vector analyzing
power in elastic Nd scattering and 
 for the neutron-deuteron (nd) 
 breakup cross sections 
in neutron-neutron (nn) quasi-free-scattering (QFS) and
symmetric-space-star (SST) geometries \cite{ref1}. Since both these
configurations depend predominantly on the S-wave nucleon-nucleon (NN) force
components, these cross section discrepancies have 
  serious  consequences 
 for the nn $^1S_0$ force component. 
 A stronger $^1S_0$ nn
force is required to bring theory and nn QFS data to agreement.
 The increased strength of the $^1S_0$ nn interaction could make the nn system
 bound.

At energies above 
$\approx 100$~MeV current 3NF's only partially improve
the description of cross section data and the remaining differences  
 indicate the possibility of relativistic
effects.  
We extended our relativistic formulation of 3N Faddeev
equations to include also 3NF \cite{ref2}. 
 New results  show that  relativistic effects based
on relativistic kinematics and boost effects of the NN 
force play an important role in building up the magnitude of
 3NF effects.  

One of the reasons for the above disagreements could be  a lack of consistency 
between 2N and 3N forces used or/and
omission of important  terms in the applied 3NF.
 The Chiral Effective Field Theory approach
 provides consistent 2N and 3N forces. 
 Recently  the chiral 3NF  at  N$^3$LO was derived \cite{ref3}, \cite{ref4}. 
At this order 3NF consists of long range parts with the
2$\pi$-exchange, 1$\pi$-2$\pi$ and ring terms \cite{ref3} and a 
short-range contributions  
2$\pi$-contact and relativistic corrections
of order 1/m \cite{ref4}. 
This is supplemented by 1$\pi$- and 3N- contact terms.
 First results obtained with that N$^3$LO force without relativistic 
 corrections  show that it does not provide explanation for the 
  low energy $A_y$ puzzle.

\setlength{\bibsep}{0.0em}
\begin{thebibliographynotitle}{99}
\bibitem{ref1} H.~Wita{\l}a and W.~Gl\"ockle,
         Phys. Rev. C83 (2011) 034004.
\bibitem{ref2} H.~Wita{\l}a {\it et al.},  Phys. Rev. C83 (2011) 044001.
\bibitem{ref3} V.~Bernard {\it et al.}, Phys. Rev. C77 (2008) 064004.
\bibitem{ref4} V.~Bernard {\it et al.}, Phys. Rev. C84 (2011) 054001. 
\end{thebibliographynotitle}

\newabstract 
\begin{center}
{\large\bf Scattering states from bound state like wave functions:
       The Coulomb case}\\[0.5cm]
{\bf Alejandro Kievsky}  \\[0.3cm]
Istituto Nazionale di Fisica Nucleare, Sezione di Pisa, 56127 Pisa, Italy\\[0.3cm]
\end{center}

An application of the integral relations derived in Ref.~\cite{ref1} 
is discussed. To this end we give the generalization of the integral
relations to the case in which more than one channel is open.
\begin{eqnarray}
B_{ij}&=&-\frac{m}{\hbar^2}<\Psi_i|H-E|F_j>  \cr
\cr
A_{ij}&=&\frac{m}{\hbar^2} <\Psi_i|H-E|{\widetilde G}_j> \cr
\cr
R^{2^{nd}}&=&A^{-1}B. \; \;\;\
\label{eq:secondij}
\end{eqnarray}
with $R^{2^{nd}}$ the second order estimate of the scattering matrix
whose eigenvalues are the phase shifts and the indices $(i,j)$ indicate
the different asymptotic configurations accessible at the specific energy
under consideration. Consider $p-d$ scattering at $E_{lab}=3$ MeV
in $J=1/2^+$ state, the scattering matrix is a $2\times 2$ matrix. 
Using the AV18 potential,
phase-shift and mixing parameters calculated using the PHH expansion,
are given in Table~\ref{tab:irtab2}. It is possible to solve an
equivalent problem with a screened Coulomb potential 
and then use the integral relations to extract
the scattering matrix corresponding to the unscreened problem~\cite{ref2}. 
This has been done using Eq.~\ref{eq:secondij} and the results are given in
Table~\ref{tab:irtab2} using $r_{sc}=50$ fm and $n_{sc}=5$~\cite{ref3}.
We observe a complete agreement between the two procedures.

\begin{table}[h]
\caption{Phase-shift and mixing parameters for $p-d$ scattering
at $E_{lab}=3$ MeV using the AV18 potential. Results using the
PHH expansion (second column) and using the integral relations
(last column).}
\begin{tabular}{lcc}
\hline
              & $p-d$         & Int.Rel.       \\
\hline
  $ ^4D_{1/2}$&$-3.563^\circ$ & $-3.562^\circ $ \\
  $ ^2S_{1/2}$&$-32.12^\circ$ & $-32.12^\circ $ \\
 $\eta_{1/2+}$&$1.100^\circ $ & $ 1.101^\circ $ \\
\hline
\end{tabular}
\label{tab:irtab2}
\end{table}

\setlength{\bibsep}{0.0em}
\begin{thebibliographynotitle}{99}
\bibitem{ref1} P. Barletta, C. Romero-Redondo, A. Kievsky, M. Viviani, and E. Garrido
    Phys. Rev. Lett. 103, 090402 (2009).
\bibitem{ref2} A. Kievsky, M. Viviani, P. Barletta, C. Romero-Redondo, and E. Garrido
    Phys. Rev. C81, 034002 (2010).
\bibitem{ref3} A.Kievsky, EPJ Web of Conferences 3, 01002 (2010).
\end{thebibliographynotitle}

\newabstract 
\begin{center}
{\large\bf Story Behind the Launch of Deuteron-Proton Scattering Experiments at
  RIKEN; How It All Started. 
}\\[0.5cm]
{\bf Hideyuki Sakai}  \\[0.3cm]
RIKEN Nishina Center,\\
Wako, Saitama 351-0198, Japan\\[0.3cm] 
\end{center}

It all started with the construction of the polarized ion source(PIS) 
for deuteron. 

Around 1990, we decided to pursue spin-isospin physics by using a
polarized deuteron beam.
The research topics considered at the time were
the ($\vec{d},^2$He) reaction for the $\beta^+$ Gamow-Teller(GT) response 
study or  
the ($\vec{d},\vec{d'}$) reaction for the double GT giant resonance search. 
To this end we proposed the construction of PIS at the RIKEN Ring
Cyclotron facility. 

The construction was finished in 1992. Before starting experiments,
absolute magnitudes of vector and tensor polarizations of 
the deuteron beam had to be measured for polarimetry.
Considering the merit and demerit, we chose 
the $\vec{d} + p$ elastic scattering.

We measured for the first time the precise complete set
of analyzing powers (A$_y$, A$_{xx}$, A$_{yy}$, A$_{xz}$) for
the $\vec{d} + p$ scattering at 270 MeV, of which I was very
proud, and we submitted the results to Physics
Letters B (PLB). However, it was turned down!
Both referees pointed out that the cross section data
were not the state-of-the-art. Indeed the data fluctuated
over scattering angle beyond the statistical errors.
This was because we were interested in the deuteron polarimetry
and did not pay much attention to the cross sections.
According to the suggestion
of the PLB Editor, we remeasured the cross sections with improved
equipment, to reduce the systematic errors. With the new 
data the paper was accepted and published \cite{sakamoto}.
This was the beginning
of our involvement in the three-nucleon-force (3NF) study.

Soon after, rigorous Faddeev calculations with 3NF became
available from the Bochum group \cite{witala}
and naturally the collaboration between Bochum group and us followed.
The first outcome of the collaboration appeared in \cite{sakai},
where the 3NF effects were clearly shown in the angular region
with the minimum cross section. 

Since then, extremely fruitful collaborations have been developed.
Our recent activities are described at this conference
by Kimiko Sekiguchi.

We would like to thank Walter Gr\"ockle for his continuous support and 
encouragement. Without them, we would not have achieved such excellent results.
\setlength{\bibsep}{0.0em}
\begin{thebibliographynotitle}{99}
\bibitem{sakamoto} Sakamoto, Sakai {\it et al.}, Phys. Lett. {\bf B 367} (1996) 60.
\bibitem{witala} Wita{\l}a {\it et al.} Phys. Rev. Lett. {\bf 81} (1998) 1183.
\bibitem{sakai} Sakai {\it et al.} Phys. Rev. Lett. {\bf 84} (2000) 5288.
\end{thebibliographynotitle}

\newabstract 
\begin{center}
{\large\bf Spin Physics and ``Polarized Fusion''}\\[0.5cm]
{\bf Hans Paetz gen.~Schieck}  \\[0.3cm]
Institut f\"ur Kernphysik, Universit\"at zu K\"oln,\\
50937 K\"oln, Germany\\[0.3cm]
\end{center}
Low-energy fusion reactions, especially the four- and five-nucleon
reactions, are not only important for big-bang nucleosynthesis, but
are the main reactions considered for energy production in future
reactors, either with magnetically (TOKOMAKS such as ITER) or
inertially confined plasmas (e.g. laser facilities such as NIF). The
primary research goal is to reach energetic break-even, before useful
reactors can be implemented, but spin-polarized fuel might speed-up
progress.  

The use of polarized particles as fusion fuel has long been considered
\cite{Kul82} and it is undisputed that for the $\mathrm{d} +
\mathrm{^3H} \to \mathrm{n} +  \mathrm{^4He}$ + 17.58 MeV (1) and $
\mathrm{d} + \mathrm{^3He} \to \mathrm{p}  + \mathrm{^4He}$ + 18.34
MeV (2) reactions the reaction rates can be enhanced by up to a factor
1.5 by polarizing the reacting nuclei. In addition, the polarization
can be used to control particle emission directions and thus save
structure materials. For a number of reasons the neutron-less reaction
(2) is preferred but needs higher incident energy and is accompanied
by neutrons from the $\mathrm{d} + \mathrm{d} \to \mathrm{n} +
\mathrm{^3He}$ + 3.268 MeV (3) reaction. This and $\mathrm{d} +
\mathrm{d} \to \mathrm{p} + \mathrm{^3H}$ + 4.033 MeV (4) are
energy-producing reactions in their own right, but are less efficient.  

The proposal of suppressing these DD neutrons by polarizing the
deuterons in the S=2 (quintet) state has incited many conflicting
theoretical predictions as well as attempts to parametrize the
existing experimental data, see e.g. \cite{Lem93} whereas a
spin-correlation experiment has never been attempted \cite{Pae10}. It
is clear that -- due to the very low energies (10 - 100 keV) -- such
an experiment is difficult and lengthy and requires sophisticated
spin-polarization techniques, but is being planned by the {\it
  PolFusion} collaboration \cite{Gat12}, \cite{Pae12}. The reaction mechanism
of the two reactions is very complicated and requires up to 16 complex
T-matrix elements (S-, P-, and D-waves are important) and is therefore
also interesting in itself. The low-energy data situation is rather
poor, and better theoretical predictions for all (polarization)
observables are urgently needed. 

\setlength{\bibsep}{0.0em}
\begin{thebibliographynotitle}{99}
\bibitem{Kul82} R.M.~Kulsrud {\it et al.}, Phys.~Rev.~Lett.~{\bf 49}, 1248 (1982).
\bibitem{Pae10} H.~Paetz gen.~Schieck, Eur.~Phys.~J.~A {\bf 44}, 321 (2010).
\bibitem{Lem93} S.~Lema{\^i}tre, H.~Paetz~gen.~Schieck, Ann.~Phys.~(Leipzig) {\bf 2}, 503 (1993).
\bibitem{Gat12}   K.~Grigoriev {\it et al.}, Proc. 19th Int. Spin
  Physics Symposium (SPIN2010), J\"ulich, J.~of Phys.~IOP Conf.~Series
  {\bf 295}, 012168 (2011). \bibitem{Pae12} H.~Paetz gen. Schieck,
  Few-Body Syst. DOI 10.1007/s00601-012-0485-0 (2012). 
\end{thebibliographynotitle}

\newabstract 
\begin{center}
{\large\bf Are New Experiments at Low Energies Really Needed to Advance Few-Body Physics?}\\[0.5cm]
{\bf Werner Tornow} \\[0.3cm]
Duke University and Triangle Universities Nuclear Laboratory,\\
Durham, NC 27708-0308, USA\\[0.3cm]
\end{center}

The reason for raising the question given in the title is the possibility that 
experimental facilities capable of producing low-energy hadron and photon 
beams will not be available anymore in the foreseeable future. Here, "low-energy"
refers to incident beams of up to about 50 MeV. Even if some existing facilities 
survive, or new ones come on-line, it will be very difficult to obtain beam time
for few-body physics experiments. Funding agencies keep saying "You have been
doing this already for too long" and they continue with the question stated in
the title. \\
In this talk I give arguments and show examples for the A=2 to A=4 few-body systems
that new experiments are needed 
\begin{enumerate}
\item[A)] If there exist observables of "special theoretical interest" for which the
   existing experimental results are contradictory.
\item[B)] If only one measurement exists for an observable of "special theoretical interest".
\item[C)] If experimental results are not available at all for observables which are of 
   "special theoretical interest".
\end{enumerate}
In general, the agreement between data and few-nucleon calculations at low energies
is very satisfactory. I consider this observation as a real success story. However,
there is still the need to definitely resolve existing discrepancies between
experimental data for a few observables, like neutron-neutron quasi-free scattering.
In addition, a few new observables should definitely be measured, for example:
\begin{enumerate}
\item[a)] The analyzing power in neutron-$^3$H elastic scattering.
\item[b)] Observables which are sensitive to the off-shell behavior of the NN interaction,
   for example, neutron-proton bremsstrahlung.
\item[c)] Some polarized spin-spin observables.
\item[d)] A full set of Wolfenstein parameters at a few energies.
\end{enumerate}
Only then it may eventually be justifiable to let Few-Body Physics at low energies
become a purely theoretical discipline.

\newabstract 
\begin{center}
{\large\bf Nuclear Dynamics from Lattice QCD}\\[0.5cm]
{\bf Silas R.~Beane}  \\[0.3cm]
Helmholtz-Institut f\"ur Strahlen- und Kernphysik (Theorie),\\
Universit\"at Bonn, D-53115 Bonn, Germany\\[0.3cm]
\end{center}

\noindent Nuclear physics is a vast field, whose phenomenology has
been explored for decades through intense experimental and theoretical
effort. However, the quantitative connection between nuclear physics
and the basic building blocks of nature, quarks and gluons, whose
interactions are encoded in QCD, are only now beginning, using lattice
QCD (LQCD) methods.  The first predictions for
nuclear physics from LQCD are for the hyperon-nucleon
interaction~\cite{Beane:2012ey}. These interactions determine, in
part, the role of the strange quark in dense matter, such as that
found in astrophysical environments.  Calculations of phase shifts,
performed at a heavy pion mass of $m_\pi\sim 390~{\rm MeV}$ have been
extrapolated to the physical pion mass using chiral effective field
theory (EFT). The interactions determined from QCD are consistent with
those extracted from hyperon-nucleon experimental data within
uncertainties.  The scattering lengths and effective ranges that
describe low-energy nucleon-nucleon scattering have been calculated in
the limit of SU(3)-flavor symmetry at the physical strange-quark mass
using LQCD~\cite{Beane:2013br}.  The values of the scattering
parameters, in the $\si$ and $\siii$ channels are, in a sense, more
natural at $m_\pi\sim 800~{\rm MeV}$ where both satisfy $a/r\sim \sim +2.0$,
than at the physical pion mass where $a^{(\si )}/r^{(\si )} \sim −8.7$
and $a^{(\siii )}/r^{(\siii )}\sim +3.1$. The relatively large size of
the deuteron compared with the range of the nuclear forces may persist
over a large range of light-quark masses, and therefore might be a
generic feature of QCD. The $\si$-channel, by contrast, is finely
tuned at the physical light-quark masses and it remains to be seen
over what range of masses this persists. Finally, the existing LQCD
data for the quark-mass dependence of nuclear binding has been used,
together with EFT methods, to show that the
cross sections for scalar-isoscalar WIMP-nucleus interactions arising
from fundamental WIMP interactions with quarks do not suffer from
significant uncertainties due to enhanced meson-exchange currents~\cite{Beane:2013kca}.

\setlength{\bibsep}{0.0em}
\begin{thebibliographynotitle}{99}

\bibitem{Beane:2012ey} 
  S.~R.~Beane, E.~Chang, S.~D.~Cohen, W.~Detmold, H.~-W.~Lin, T.~C.~Luu, K.~Orginos and A.~Parreno {\it et al.},
  Phys.\ Rev.\ Lett.\  {\bf 109}, 172001 (2012)
  [arXiv:1204.3606 [hep-lat]].

\bibitem{Beane:2013br} 
  S.~R.~Beane, E.~Chang, S.~D.~Cohen, W.~Detmold, P.~Junnarkar, H.~W.~Lin, T.~C.~Luu and K.~Orginos {\it et al.},
  arXiv:1301.5790 [hep-lat].

\bibitem{Beane:2013kca} 
  S.~R.~Beane, S.~D.~Cohen, W.~Detmold, H.~-W.~Lin and M.~J.~Savage,
  arXiv:1306.6939 [hep-ph].

\end{thebibliographynotitle}

\newabstract 
\begin{center}
{\large\bf SU(3) chiral dynamics revisited}\\[0.5cm]
{\bf V\'eronique Bernard}$^1$, S. Descotes-Genon$^2$ and G. Toucas$^2$  \\[0.3cm]
$^1$ IPN, UMR 8608,
91406 Orsay Cedex, France\\[0.3cm]
$^2$ LPT, UMR 8627, 91406 Orsay Cedex, France\\[0.3cm]
\end{center}

The analysis of the fermion determinant in terms of Dirac eigenvalues 
shows that two particularly interesting order parameters of
the spontaneous breaking of chiral symmetry, the quark condensate 
and the meson decay constant in the chiral limit, are dominated by the
infrared extremity of the Dirac spectrum. This leads to a parametric 
suppression of the three-flavor condensate namely:
\begin{equation}
\Sigma(3) <\Sigma(2) \, , \quad \quad F^2(3) <F^2(2)
\end{equation}
Interesting questions thus are:  how strong this suppression is and 
what the consequences are. To address them we have studied meson
properties such as spectrum, decay constants and $K_{\ell 3}$ form
factors on one  
hand \cite{bdt1}  and 
topological quantities \cite{bdt2}, the topological susceptibility and 
the fourth cumulant,  on the other hand in the framework of resummed $\chi PT$ 
\cite{res}.
Such a scheme reorders the chiral series  
allowing for a numerical competition between leading order and next
to leading order terms. Performing a fit to data from  two lattice collaborations
\cite{pacs}, \cite{rbc} the emerging picture
for the pattern of
chiral symmetry breaking is marked by a strong dependence of the observables
on the strange quark mass and thus a significant difference between chiral
symmetry breaking in the $N_f=2$ and $N_f=3$ chiral limits. This 
result impacts 
the chiral extrapolation of lattice data, in particular it can affect
the determination of 
the lattice spacing as discussed in \cite{bdt2}. Furthermore,     
for hierarchies of light-quark masses close to the physical situation,
the fourth cumulant has a much better sensitivity than the topological susceptibility
to the three-flavour quark condensate. Also 
a combination of the topological susceptibility and the fourth
cumulant is able to pin down the three flavour condensate in a very clean way in
the case of three degenerate quarks.

\setlength{\bibsep}{0.0em}
\begin{thebibliographynotitle}{99} 
\bibitem{bdt1} V. Bernard, S. Descotes-Genon and G. Toucas, JHEP 1101 (2011) 107.
\bibitem{bdt2} V. Bernard, S. Descotes-Genon and G. Toucas, JHEP 1206
  (2012) 051; 
JHEP 1212 (2012) 080. 
\bibitem{res} S. Descotes-Genon, N. H. Fuchs, L. Girlanda and J. Stern, Eur. Phys. J. C 34 (2004) 201.
\bibitem{pacs} S. Aoki {\it et al.} [PACS-CS Collaboration], Phys. Rev. D79 (2009) 034503.
\bibitem{rbc} C. Allton  {\it et al.} [RBC-UKQCD Collaboration], Phys. Rev. D78 (2008) 114509; P.A. Boyle  {\it et al.}, Phys. Rev. Lett. 100 (2008) 141601; 
arXiv:1004.0886.
\end{thebibliographynotitle}

\newabstract 
\begin{center}
{\large\bf Space-time picture of chiral dynamics with nucleons}\\[0.5cm]
Carlos Granados$^1$, {\bf Christian Weiss}  \\[0.3cm]
Theory Center, Jefferson Lab, Newport News, VA 23606, USA
\\[0.3cm]
\end{center}
We explore chiral dynamics with nucleons in the space--time picture
based on the partonic (or light-front) formulation of relativistic systems.
The electromagnetic form factors are expressed in terms of 
frame--independent densities of charge and magnetization in transverse 
space (see \cite{Miller:2010nz} for a review). 
The chiral component of nucleon structure is 
identified as the region of transverse distances $b = O(M_\pi^{-1})$ 
and can be studied systematically using chiral EFT methods. 
We compute the peripheral transverse charge and magnetization densities 
in the leading--order approximation and study their properties 
(large--distance behavior, heavy--baryon expansion, role of 
intermediate $\Delta$ isobars, chiral vs.\ non-chiral contributions)
\cite{Granados:2013moa}, \cite{Strikman:2010pu}.  
We demonstrate the equivalence of the Lorentz-invariant formulation 
of chiral EFT and the time-ordered formulation using 
light--front wave functions. The time--ordered formulation provides 
a simple ``mechanical'' 
interpretation of the nucleon's chiral component and explains the 
relative magnitude of the peripheral charge and magnetization 
densities. The space-time picture of chiral dynamics described here 
offers new perspectives on peripheral nucleon structure and
basic properties of the chiral expansion. It connects chiral 
dynamics with the nucleon's quark/gluon structure 
(generalized parton distributions) probed in peripheral high-energy 
scattering processes \cite{Strikman:2009bd}, \cite{Strikman:2003gz}. 
\setlength{\bibsep}{0.0em}
\begin{thebibliographynotitle}{99}
\bibitem{Miller:2010nz} 
  G.~A.~Miller,
  Ann.\ Rev.\ Nucl.\ Part.\ Sci.\  {\bf 60}, 1 (2010).
%
%
\bibitem{Granados:2013moa} 
  C.~Granados and C.~Weiss,
  arXiv:1308.1634 [hep-ph].
%
%
\bibitem{Strikman:2010pu} 
  M.~Strikman and C.~Weiss,
  Phys.\ Rev.\ C {\bf 82}, 042201 (2010).
%
%
\bibitem{Strikman:2009bd}
  M.~Strikman and C.~Weiss,
  Phys.\ Rev.\  D {\bf 80}, 114029 (2009).
%
%
\bibitem{Strikman:2003gz} 
  M.~Strikman and C.~Weiss,
  Phys.\ Rev.\ D {\bf 69}, 054012 (2004).
%
%
\end{thebibliographynotitle}

\newabstract 
\begin{center}
{\large\bf Pion production in nucleon-nucleon collisions}\\[0.5cm]
{\bf Vadim Baru}   \\[0.3cm]
 Institut f\"ur Theoretische Physik II, Ruhr-Universit\"at Bochum,\\
44780 Bochum, Germany\\  
 Institute for Theoretical and Experimental Physics,\\
 117218, B.~Cheremushkinskaya 25, Moscow, Russia
\end{center}

The proper treatment of $NN\to NN\pi$ within chiral EFT  requires taking into account 
the  intermediate momentum scale,   $p\approx \sqrt{m_{\pi} m_N}$  
with $m_{\pi} $ and $m_N$ being the pion and  nucleon  masses \cite{hanhart04}. 
Given this modification, the  application  of   chiral EFT  to  s-wave  pion production  in  the  
$pp\to d\pi^+$ channel  at  next-to-leading order  (NLO) \cite{lensky2},
revealed   good  agreement   with  experimental  data.     
In contrast, for  s-wave pion production in  the   channel  $pp\to pp\pi^0$,  
all operators at  LO  and  NLO  are   strongly  suppressed  \cite{hanhart04}.    
Therefore,  it is not a surprise that 
 the relative importance of  chiral loops with explicit nucleon and $\Delta(1232)$ degrees of freedom
  at     N$^2$LO in $pp\to pp\pi^0$ is significantly  enhanced compared to $pp\to d\pi^+$ \cite{NNLOswave}.

The study of   pion production  in $pn\to d\pi^0$   provides access to  charge symmetry breaking (CSB) phenomena
[4-6].  
CSB  in $pn\to d\pi^0$  is caused  predominantly by the strong  contribution  to the proton-neutron
mass difference \cite{filin}.  Therefore, studying CSB in $pn\to d\pi^0$
allows one to  extract separately the important low-energy parameters --  
strong and electromagnetic contributions to the proton-neutron
mass difference --  in analogy  to a recent  determination of the s-wave
$\pi N$  scattering lengths from  pionic atoms   \cite{JOB}. 
Another important application   would be to verify the
connection   provided by chiral symmetry between  pion production and  other low-energy few-nucleon
reactions. Specifically,   p-wave pion production  
can be used to pin down the $(\bar N N)^2 \pi$ low-energy constant
(LEC) [8,9].
Apart from $NN\to NN\pi$,  this LEC contributes to the 3N force, 
to electroweak processes  and 
to  reactions involving photons. 
Recent measurements of  $\vec p p\to (pp)_s \pi^0$ and  $\vec p n\to (pp)_s \pi^-$    at  COSY  
provide  the  database for  the  extraction of this LEC    \cite{ANKEAxx}. 
 
 \vspace*{-0.5cm}

\setlength{\bibsep}{0.0em}
\begin{thebibliographynotitle}{99}
 
\bibitem{hanhart04}
  C.~Hanhart,
  Phys.\ Rept.\  { 397}   (2004) 155.

\bibitem{lensky2}
  V.~Lensky {\it et al.}, 
  Eur.\ Phys.\ J.\ A\ { 27} (2006) 37.

\bibitem{NNLOswave}
A.~A.~Filin {\it et al.},
  Phys.\ Rev.\ C { 85} (2012) 054001;   
arXiv:1307.6187  (2013).

\bibitem{jouni}
U.~van Kolck  et  al., 
\newblock {\em Phys. Lett.}, B {493}, 65 (2000).

\bibitem{filin}
  A.~Filin {\it et al.},
  Phys.\ Lett.\ B { 681} (2009) 423.

\bibitem{Opper}
A.~K. Opper et~al.
\newblock {\em Phys. Rev. Lett.}, {91}, 212302 (2003).

\bibitem{JOB}
 V.~Baru {\it et al.}, 
  Phys.\ Lett.\ B {694} (2011) 473;
  Nucl.\ Phys.\ A {872} (2011) 69,

\bibitem{ch3body}
  C.~Hanhart  {\it et al.},  
  Phys.\ Rev.\ Lett.\ \ {85} (2000) 2905.
  
\bibitem{newpwave}
  V.~Baru {\it et al.}, 
  Phys.\ Rev.\ C\ {80} (2009) 044003.

\bibitem{ANKEAxx}
  S.~Dymov {\it et al.}  [ANKE Collaboration],
  arXiv:1304.3678 [nucl-ex].
  
\end{thebibliographynotitle}

\newabstract 
\begin{center}
{\large\bf Issues of renormalization in EFT for the NN system}\\[0.5cm]
Evgeny Epelbaum$^1$, and {\bf Jambul Gegelia}$^{1,2}$  \\[0.3cm]
$^1$Institut f\"ur Theoretische Physik II, Ruhr-Universit\"at Bochum,\\
44780 Bochum, Germany\\[0.3cm]
$^2$Tbilisi State  University,  
 0186 Tbilisi,
 Georgia  \\[0.3cm]
\end{center}

In calculations of physical quantities in low-energy effective field
theory (EFT) of few nucleons \cite{Weinberg1}, 
one usually interchanges the order of calculating quantum corrections and the non-relativistic expansion.   
Generally, to perform a proper non-relativistic expansion, one has to
first evaluate the observable of interest based on a Lorentz-invariant effective Lagrangian.
The calculation of quantum corrections (i.e.~loop diagrams) requires
regularization  ($\Lambda$) and renormalization. Resulting finite
physical quantities can then be expanded in inverse powers of the nucleon mass ($m$).
On the other hand, in non-relativistic formulations of EFT, one first 
expands the Lorentz-invariant effective Lagrangian in powers of
$1/m$. 
The calculation of quantum corrections based on the 
$1/m$-expanded Lagrangian again requires 
regularization and renormalization and leads to 
renormalized quantities being represented  as series in $1/m$.
Generally, the expansion in $1/m$ and calculation of quantum
corrections are non-commutative. However,
the difference (''error'') can be compensated by adding terms to
the $1/m$-expanded EFT Lagrangian \cite{Gegelia:1999gf}.
Due to non-commutativity of $ 1/m$ and $1/\Lambda$ expansions in loop
integrals and nonperturbative nature of the problem at hand, an infinite number of compensating terms are 
to be taken into account in the NN sector already at leading order (LO). At
least in some cases, these contributions play a crucial role and hence
cannot be dropped \cite{Epelbaum:2009sd}. 
Possible solutions to this problem include   
keeping $\Lambda \lesssim m$ (see e.g. \cite{Epelbaum:2008ga})
and 
using the original Lorentz invariant Lagrangian without interchanging the $1/m$-expansion and the calculation of loop contributions.
The last approach has been formulated and applied at LO in Refs.~[5-7]. 

This work was supported by
the DFG (GE 2218/2-1) and
the Georgian Shota Rustaveli National Science Foundation (grant 11/31).

\setlength{\bibsep}{0.0em}
\begin{thebibliographynotitle}{99}
\bibitem{Weinberg1} S.~Weinberg, Phys. Lett. B251 (1990) 288.
\bibitem{Epelbaum:2008ga}
  E.~Epelbaum {\it et al.}, 
  Rev. Mod.\ Phys.  {\bf 81}, 1773 (2009).
\bibitem{Gegelia:1999gf}
  J.~Gegelia and G.~Japaridze,
  Phys.\ Rev.\ D {\bf 60}, 114038 (1999).
\bibitem{Epelbaum:2009sd}
  E.~Epelbaum and J.~Gegelia,
  Eur.\ Phys.\ J.\ A {\bf 41}, 341 (2009).
\bibitem{Epelbaum:2012ua}
  E.~Epelbaum and J.~Gegelia,
  Phys.\ Lett.\ B {\bf 716}, 338 (2012).
\bibitem{Epelbaum:2012cv}
  E.~Epelbaum and J.~Gegelia,
  arXiv:1210.3964 [nucl-th].
\bibitem{Epelbaum:2013ij}
  E.~Epelbaum and J.~Gegelia,
  arXiv:1301.6134 [nucl-th].
\end{thebibliographynotitle}

\newabstract 
\begin{center}
{\large \textbf{Overview and Latest News }}\\[0pt]
{\large \textbf{from Nuclear Lattice Effective Field Theory}}\\[0.5cm]
\textbf{Dean Lee} [for the Nuclear Lattice EFT\ Collaboration]\\[0.3cm]
Department of Physics, North Carolina State University, Box 8202,%
\\[0pt]
Raleigh, NC 27695, USA\\[0.3cm]
\end{center}

This talk begins with an overview of the theoretical methods and numerical
algorithms used in nuclear lattice effective field theory. \ This includes a
description of chiral effective field theory regulated on the lattice \cite%
{Borasoy:2006qn}, \cite{Borasoy:2007vi}, \cite{Epelbaum:2009zsa}, scattering phase shifts on
the lattice using spherical wall boundaries \cite{Borasoy:2007vy}, and Monte
Carlo simulations using Euclidean time projection and auxiliary fields \cite%
{Lee:2008fa}.

The remainder of the talk presents recent results on the spectrum of
carbon-12 \cite{Epelbaum:2009pd}, the structure and rotations of the Hoyle
state \cite{Epelbaum:2011md}, \cite{Epelbaum:2012qn}, preliminary new results on
nuclei up to $A=28$, and the spectrum and structure of oxygen-16.

\setlength{\bibsep}{0.0em}

\begin{thebibliographynotitle}{9}
\bibitem{Borasoy:2006qn} B.~Borasoy, E.~Epelbaum, H.~Krebs, D.~Lee and
U.~-G.~Mei\ss ner, 
Eur.\ Phys.\ J.\ A \textbf{31}, 105 (2007).

\bibitem{Borasoy:2007vi} B.~Borasoy, E.~Epelbaum, H.~Krebs, D.~Lee and
U.~-G.~Mei\ss ner, 
Eur.\ Phys.\ J.\ A \textbf{35}, 343 (2008).

\bibitem{Epelbaum:2009zsa} E.~Epelbaum, H.~Krebs, D.~Lee and U.~-G.~Mei\ss %
ner, 
Eur.\ Phys.\ J.\ A \textbf{41}, 125 (2009).

\bibitem{Borasoy:2007vy} B.~Borasoy, E.~Epelbaum, H.~Krebs, D.~Lee and
U.~-G.~Mei\ss ner, 
Eur.\ Phys.\ J.\ A \textbf{34}, 185 (2007).

\bibitem{Lee:2008fa} D.~Lee, 
Prog.\ Part.\ Nucl.\ Phys.\ \textbf{63}, 117 (2009).

\bibitem{Epelbaum:2009pd} E.~Epelbaum, H.~Krebs, D.~Lee and U.~-G.~Mei\ss %
ner, 
Phys.\ Rev.\ Lett.\ \textbf{104}, 142501 (2010).

\bibitem{Epelbaum:2011md} E.~Epelbaum, H.~Krebs, D.~Lee and U.~-G.~Mei\ss %
ner, 
Phys.\ Rev.\ Lett.\ \textbf{106}, 192501 (2011).

\bibitem{Epelbaum:2012qn} E.~Epelbaum, H.~Krebs, T.~A.~L\"{a}hde, D.~Lee and
U.~-G.~Mei\ss ner, 
Phys.\ Rev.\ Lett.\ \textbf{109}, 252501 (2012).
\end{thebibliographynotitle}

\newabstract 
\begin{center}
{\large\bf Life on Earth -- an accident?}\\[0.5cm]
{\bf Ulf-G. Mei{\ss}ner}$^{1,2}$ [for the NLEFT collaboration]  \\[0.3cm]
$^1$HISKP and BCTP, Universit\"at Bonn,
53115 Bonn, Germany\\[0.3cm]
$^2$IKP, IAS and JCHP, Forschungszentrum J\"ulich, \\
52425 J\"ulich, Germany \\[0.3cm]
\end{center}

The Hoyle state is considered a prime example for the {\em anthropic
  principle}, that states that the fundamental parameters of all
interactions are constrained by the requirement that carbon-oxygen based 
life could have been formed. Carbon is formed in hot stars via the
triple-alpha process and this requires an $0^+$ excited state in $^{12}$C
close to the 3$\alpha$ threshold, the so-called Hoyle state. This state
could be described {\em ab initio} for the first time using nuclear
lattice simulations \cite{Epelbaum:2011md}. In this framework, one can
address the question: how much  fine-tuning in the fundamental parameters of
QCD+QED (the part of the Standard Model underlying nuclear physics) is
allowed to still have a sufficient production of the life-essential elements?
This requires the knowledge of the quark mass dependence of the nuclear
forces as  addressed to next-to-next-to-leading order in chiral EFT 
in Ref.~\cite{Berengut:2013nh}. Based on this,
one can study the ground and excited state energies of the nuclei appearing in
the   3$\alpha$-process, see Refs.~\cite{Epelbaum:2012iu}, \cite{Epelbaum:2013wla}.
We find strong evidence that the physics of the $3\alpha$-process is 
driven by $\alpha$-clustering, and that shifts in the light quark mass 
at the $\simeq 2 - 3 \%$ level are unlikely to be detrimental to the
development  of life. Tolerance against much larger changes cannot be ruled 
out at present, given the relatively limited knowledge of the quark mass 
dependence of the two-nucleon S-wave scattering parameters. Lattice QCD is 
expected to provide refined estimates of the scattering parameters in the future.
Further, variations of the fine-structure constant $\alpha_{\rm EM}$ up to
$\pm 2.5\%$ are consistent with the requirement of sufficient carbon and
oxygen production in stars.

\setlength{\bibsep}{0.0em}
\begin{thebibliographynotitle}{99}

\bibitem{Epelbaum:2011md}
  E.~Epelbaum, H.~Krebs, D.~Lee and U.-G.~Mei{\ss}ner,
  Phys.\ Rev.\ Lett.\  {\bf 106} (2011) 192501
  [arXiv:1101.2547 [nucl-th]].

\bibitem{Berengut:2013nh}
  J.~C.~Berengut,
  E.~Epelbaum, V.~V.~Flambaum, C.~Hanhart, U.-G.~Mei{\ss}ner, J.~Nebreda and J.~R.~Pelaez,
  Phys.\ Rev.\ D {\bf 87} (2013) 085018
  [arXiv:1301.1738 [nucl-th]].

\bibitem{Epelbaum:2012iu}
  E.~Epelbaum, H.~Krebs, T.~A.~L\"ahde, D.~Lee and U.-G.~Mei{\ss}ner,
  Phys.\ Rev.\ Lett.\  {\bf 110} (2013) 112502
  [arXiv:1212.4181 [nucl-th]].

\bibitem{Epelbaum:2013wla}
  E.~Epelbaum, H.~Krebs, T.~A.~L\"ahde, D.~Lee and U.-G.~Mei{\ss}ner,
  Eur. Phys. J. A {\bf 49}:82 (2013)
  [arXiv:1303.4856 [nucl-th]].
\end{thebibliographynotitle}

\newabstract 
\begin{center}
{\large\bf Nuclear Structure and Reactions in Lattice Effective Field Theory}\\[0.5cm]
{\bf Gautam Rupak}  \\[0.3cm]
Department of Physics \& Astronomy, Mississippi State University\\
Mississippi State, Mississippi 39762, USA \\[0.3cm]
\end{center}

Nuclear structure and reactions play an important role in low-energy
nuclear astrophysics where constraints on fundamental physics can be
placed.  
 There is also renewed experimental interest in properties of nuclei
 that populate areas in the nuclear chart far from the valley of
 stability.   
Nuclear theory is needed for extrapolation of experimental data  to
the low-energy regime relevant to nuclear astrophysics.  \emph{Ab
  initio} calculation becomes important when parameters necessary for
constraining the nuclear theory are not know experimentally.  This is
the case~\cite{RupakA}, for example, for  \nLi~ and \pBe.  
Thus there is a physics need for ab initio calculations. Further,
microscopic calculations based on chiral perturbation theory
($\chi$PT) can act as the bridge between nuclear properties and first
principle Quantum Chromodynamic (QCD) calculations in lattice gauge
theories.  Lattice  effective field theory (EFT)calculations are
particularly attractive as it systematically combines  $\chi$PT rooted
in QCD with the powerful numerical lattice
methods~\cite{LeeA}. However, there has been no general method for
calculating reactions on the lattice. 

I present a general method for calculating radiative capture reactions
$a(b,\gamma)c$ where $a$, $b$, $c$ are generic
nuclei~\cite{RupakB}. The basic idea involves calculating an effective
two-body Hamiltonian for scattering the nuclei $a$ and $b$ using an
adiabatic Euclidean time projection, and using this Hamiltonian to
calculate the capture reactions on the lattice. We test the method by
calculating the adiabatic Hamiltonian for fermion-dimer and quartet
channel neutron-deuteron scattering.  Calculation of the capture
reaction using a two-body Hamiltonian on the lattice is demonstrated
by considering   $p(n,\gamma)d$. We use the two-point retarded Green's
function to calculate the capture reaction where we introduce an
infrared regulator to systematically suppress the finite volume
corrections on the lattice.  Finally, I present some new calculations
of neutron matter equation of state  
that updates the results from Ref.~\cite{EpelbaumA}
using an improved lattice EFT Hamiltonian. 

\setlength{\bibsep}{0.0em}
\begin{thebibliographynotitle}{99}
\bibitem{RupakA} G. Rupak and R. Higa,  Phys. Rev. Lett. 106 (2011) 222501. 
\bibitem{LeeA} D. Lee, Prog. Part. Nucl. Phys. 63 (2009) 117.
\bibitem{RupakB} G. Rupak and D. Lee, Phys. Rev. Lett. 111 (2013) 032502. 
\bibitem{EpelbaumA} E. Epelbaum {\it et al.},  Eur. Phys. J. A 40 (2009) 199.
\end{thebibliographynotitle}

\newabstract 
\begin{center}
{\large\bf Evolving perspectives on the origins of nuclear structure}\\[0.5cm]
{\bf James P. Vary}  \\[0.3cm]
Department of Physics and Astronomy, Iowa State University,\\
Ames, IA     50011, USA\\[0.3cm]
\end{center}

The overarching problem of nuclear theory is to use all the
fundamental interactions of the  Standard Model to describe and
predict nuclear phenomena.  The ``ab initio'' approach to nuclear
structure, in which one solves the nuclear many-body problem with
interactions based on Chiral Effective Field Theory (EFT), is a key
step in this direction and it is emerging as a validated and
predictive theory with quantified uncertainties. 

Among the many recent successes to mention, the successful description
of the suppressed beta decay of 14C as rooted in the role of the
chiral three-nucleon interaction stands as a particularly interesting
achievement \cite{Maris:2011as}.  Historically, many-body theory with
realistic nucleon-nucleon interactions alone had been unsuccessful in
achieving the required suppression.  It emerged that this decay is
sensitive to subtle changes in the spin-dependent components of the
nuclear Hamiltonian that were successfully described only with the
addition of the three-nucleon interaction.  Additional advances in
both structure and reactions have been reviewed recently
\cite{Barrett2013131}, \cite{Maris2013}. 


Progress to date allows us to demonstrate certain limitations of
currently available Hamiltonians from Chiral EFT.  This motivates a
new generation of efforts to derive and implement improved
Hamiltonians.  A large-scale collaboration (LENPIC) is getting
underway with these goals in mind. 

Many challenges for microscopic many-body theory of nuclei remain
including the need for a detailed description of cluster structure in
nuclei as well as other collective motions.  We also aim to extend ab
initio microscopic many-body theory to heavier nuclei than those
currently addressed. We have entered an era where large-scale
calculations are required as well as continued collaborations between
physicists, applied mathematicians and computer scientists. 

This work is supported by the US DOE SciDAC program through the NUCLEI
collaboration, by the US DOE Grants No. DE-SC-0008485 (SciDAC/NUCLEI)
and No. DE-FG02 87ER40371 as well as by the US NSF Grant No. 0904782.   
\setlength{\bibsep}{0.0em}
\begin{thebibliographynotitle}{99}
\bibitem{Maris:2011as} 
  P.~Maris, J.~P.~Vary, P.~Navr\'atil, W.~E.~Ormand, H.~Nam and D.~J.~Dean,
  Phys.\ Rev.\ Lett.\  {\bf 106}, 202502 (2011)
  [arXiv:1101.5124 [nucl-th]].
  
  \bibitem{Barrett2013131}
B.~R. Barrett, P.~Navr\'atil and J.~P. Vary, {\em Prog. Part. Nucl. Phys.} {\bf
  69}  (2013) 131.

\bibitem{Maris2013}
P.~Maris and J.~P. Vary, 
{\em Int. J. Mod. Phys. E.}, (in press).

\end{thebibliographynotitle}

\newabstract 
\begin{center}
{\large\bf New Horizons in Ab Initio Nuclear Structure Theory}\\[0.5cm]
{\bf Robert Roth}  \\[0.3cm]
Institut f\"ur Kernphysik, Technische Universit\"at Darmstadt, \\
Schlossgartenstr. 2, 64289 Darmstadt, Germany \\[0.3cm]
\end{center}

Low-energy nuclear theory has entered an era of ab initio nuclear
structure and reaction calculations based on input from QCD. One of
the most promising paths from QCD to nuclear observables employs
Hamiltonians constructed within chiral effective field theory (EFT) as
starting point for precise ab initio studies. However, the full
inclusion of chiral two- plus three-nucleon (NN+3N) interactions in
exact and approximate many-body calculations beyond the few-body
domain poses a challenge.  

Our recent breakthroughs enable ab initio calculations for ground
states and spectra of nuclei throughout the p- and lower sd-shell with
full 3N interactions using consistent Similarity Renormalization Group
(SRG) transformations and the Importance-Truncated No-Core Shell Model
(IT-NCSM) \cite{1}, \cite{2}. In this ab initio framework we are studying,
e.g., the ground-state properties and spectroscopy along the Carbon
and Oxygen isotopic chains and demonstrate the predictive power of
chiral Hamiltonians \cite{2}. The same NN+3N Hamiltonians can be
applied for the ab initio description of medium-mass nuclei, e.g. in
Coupled Cluster Theory  \cite{3}, \cite{4} or in the In-Medium SRG
\cite{5}, \cite{2}. These calculations clearly demonstrate that the frontier
of ab initio nuclear structure theory is rapidly moving towards
heavier nuclei, which were completely out of reach a few years
ago. Another new direction is the ab initio description of p-shell
hypernuclei, which was the domain of phenomenological calculations so
far. With the advent of hyperon-nucleon interactions from chiral EFT
and the advances in many-body methods ab initio calculations of ground
states and spectroscopy of single-$\Lambda$ hypernuclei are now
possible throughout and beyond the p-shell.  

As these developments show, the horizon for ab initio nuclear
structure theory has changed dramatically over the past few years,
offering exciting perspectives for QCD-based nuclear structure physics
in the coming years. 

Supported by the DFG through SFB 634, by the Helmholtz International
Center for FAIR, and by the BMBF through BMBF-FSP 302 and 06DA7047I.

\setlength{\bibsep}{0.0em}
\begin{thebibliographynotitle}{99}

\bibitem{1} R. Roth, et al.; Phys. Rev. Lett. 107, 072501 (2011).
\bibitem{2} H. Hergert, et al.; Phys. Rev. Lett. 110, 242501 (2013).
\bibitem{3} S. Binder, et al.; Phys. Rev. C 87, 021303(R) (2013).
\bibitem{4} R. Roth, et al.; Phys. Rev. Lett 109, 052501 (2012).
\bibitem{5} H. Hergert, et al.; Phys. Rev. C 87, 034307 (2013).
 
\end{thebibliographynotitle}

\newabstract 
\begin{center}
{\large\bf {\boldmath $Ab~initio$} many-body calculations of light-ion reactions}\\[0.5cm]
{\bf C. Romero-Redondo}$^1$, P. Navr\'atil$^1$, S. Quaglioni$^2$,
G. Hupin$^2$,\\ J. Langhammer$^3$, A. Calci$^3$ and R. Roth$^3$
  \\[0.3cm]
$^1$TRIUMF, 4004 Wesbrook Mall, Vancouver, BC V6T 2A3, Canada\\
$^2$Lawrence Livermore National Laboratory, P.O. Box 808, L-414, Livermore, California 94551, USA\\
$^3$Institut f\"ur Kernphysik, Technische Universit\"at Darmstadt, D-64289 Darmstadt, Germany\\[0.3cm]
\end{center}

The $ab~initio$ no-core shell model/resonating group method (NCSM/RGM)
introduced in Refs. \cite{bib1}, \cite{bib2} is a technique capable of 
describing  both structure and reactions in light nuclear systems. This approach 
combines  a microscopic cluster technique with the use of realistic 
inter-nucleon
interactions and a consistent microscopic description of the nucleon clusters.

The method has been introduced in detail for two-body clusters and has 
been shown to work efficiently in different systems
[1-4]. In this work we discuss recent advances
of the method which include 
its coupling with the NCSM into a new approach called
no-core shell model with continuum (NCSMC) with
results for $^7$He resonances \cite{baroni}. We also present the first results after the inclusion 
of chiral three-nucleon forces in the calculations and its effect in the 
nucleon-$^4$He scattering phase shifts \cite{hup13}.
Finally,  we introduce three-body cluster 
configurations 
and provide, for the first time within an
$ab~initio$ framework, the correct asymptotic
behaviour for the three-cluster wave functions. 
We present the results obtained for  $^6$He within a 
$^4$He(g.s.)+n+n basis for the ground and continuum states \cite{qua13}.


\paragraph{Acknowledgments:} Prepared in part by LLNL under Contract DE-AC52-07NA27344. Support from the
NSERC Grant No. 401945-2011, U.S. DOE/SC/NP (Work Proposal No. SCW1158),
 the DFG through SFB 634, 
 the Helmholtz International Center for FAIR 
and the BMBF(06DA7074I)
is acknowledge.

\setlength{\bibsep}{0.0em}
\begin{thebibliographynotitle}{99}
\bibitem{bib1} S. Quaglioni and P. Navr\'atil, Phys. Rev. Lett. 101 (2008) 092501.
\bibitem{bib2} S. Quaglioni and P. Navr\'atil, Phys. Rev. C 79 (2009) 044606.

\bibitem{bib4}  P. Navr\'atil and S. Quaglioni, Phys. Rev. Lett. 108 (2012) 042503.
\bibitem{nav11} P. Navr\'atil, R. Roth,  and S. Quaglioni, Phys. Lett. B 704 (2011) 379.

\bibitem{baroni} S. Baroni, P. Navr\'atil, S. Quaglioni, Phys. Rev. C 87 (2013) 034326.
\bibitem{hup13} G. Hupin, J. Langhammer, P. Navr\'atil, S. Quaglioni, A. Calci, R. Roth, in preparation (2013).
\bibitem{qua13} S. Quaglioni, C. Romero-Redondo and P. Navr\'atil, in preparation (2013).

\end{thebibliographynotitle}

\newabstract 
\begin{center}
{\large\bf Neutron rich matter from chiral effective field theory interactions}\\[0.5cm]
{\bf Kai Hebeler}  \\[0.3cm]
Institut f\"ur Kernphysik,
Technische Universit\"at Darmstadt,64289 Darmstadt, Germany\\
ExtreMe Matter Institute EMMI, GSI Helmholtzzentrum f\"ur Schwerionenforschung GmbH, 64291 Darmstadt, Germany
\\[0.3cm]
\end{center}

Chiral effective field theory (EFT) offers a systematic expansion of nuclear forces well suited to meet
the calculational challenges of neutron-rich matter, which span the extremes from universal properties at low-densities to 
the dense matter in neutron stars. The development of novel Renormalization Group (RG) methods make it possible to 
study various properties of matter within many-body perturbation theory~\cite{PNM}. Very recently these results have 
been validated for the first time using Quantum Monte Carlo calculations based on chiral EFT interactions~\cite{QMC}. 
Our current efforts aim at improving the treatment of chiral three-nucleon (3N) forces by employing interactions which 
have been evolved consistently within the Similarity RG~\cite{3N_evolution}.

Based on our microscopic neutron matter results up to densities around nuclear saturation, we were able to derive 
model-independent constraints on the nuclear equation of state at higher densities using only constraints from 
neutron star mass observations and causality~\cite{NS}.

Neutron matter also provides a powerful laboratory for testing chiral
EFT power counting at relevant nuclear densities, since  
only long-range 3N forces contribute at next-to-next-to-leading order
(N$^2$LO)~\cite{PNM} and there are no new parameters for 3N and
four-nucleon interactions at next-to-next-to-next-to-leading order
(N$^3$LO). In Ref.~\cite{N3LO} we presented the first complete N$^3$LO
calculation of the neutron matter energy, including contributions from
3N and 4N forces. We find large contributions from 3N forces at
N$^3$LO, which indicates that a chiral EFT  
with explicit delta degrees of freedom might be more efficient.

\setlength{\bibsep}{0.0em}
\begin{thebibliographynotitle}{99}

\bibitem{PNM}
K.\ Hebeler and A.\ Schwenk, Phys.\ Rev.\ C82 (2010) 014314.

\bibitem{QMC}
A. Gezerlis {\it el al.}, arXiv:1303.6243 [nucl-th].

\bibitem{3N_evolution}
K.\ Hebeler, Phys.\ Rev.\ C85 (2012) 021002(R); K.\ Hebeler and R.\ J.\ Furnstahl, Phys.\ Rev.\ C87 (2013) 031302(R).

\bibitem{NS}
K.\ Hebeler {\it et al.}, Phys.\ Rev.\ Lett.\ 105 (2010) 161102; K.\ Hebeler {\it et al.}, Astrophys. J., 773 (2013) 11.

\bibitem{N3LO}
I.\ Tews {\it et al.}, Phys.\ Rev.\ Lett. 110 (2013) 032504; T.\ Kr\"uger {\it et al.}, arXiv:1304.2212 [nucl-th].


\end{thebibliographynotitle}

\newabstract 
\begin{center}
{\large\bf Universal Properties of Halo Nuclei}\\[0.5cm]
{\bf H.-W. Hammer}  \\[0.3cm]
Helmholtz-Institut f\"ur Strahlen-und Kernphysik and
Bethe Center for Theoretical Physics, Universit\"at Bonn, 53115 Bonn, 
Germany\\[0.3cm]
\end{center}

Halo Effective Field Theory (EFT) exploits the separation of scales
between a tightly bound core and loosely bound valence 
nucleons in halo nuclei.  Low-energy observables are described in a controlled
expansion in the ratio $R_{\rm core}/R_{\rm halo}$
of the scales characterizing core and
halo~\cite{Bertulani-02}.  While ab initio
approaches try to predict nuclear observables from a 
fundamental nucleon-nucleon interaction, halo EFT essentially provides 
model-independent relations between different nuclear observables.
Coulomb effects and electromagnetic currents can be included in a 
straightforward way.

In this talk, we discuss recent results on the charge and matter
form factors of halo nuclei.
We review the universal properties
and structure of one-neutron ($1n$) and $2n$ halo nuclei, focusing on
their form factors and radii \cite{Canham:2008jd}.
In particular, we highlight recent work by Acharya 
et al.~\cite{Acharya:2013aea} on the implications of a 
recent matter radius measurement for the binding energy 
and existence of excited Efimov states in $^{22}$C. 
Moreover, we discuss the extension of the electromagnetic
structure calculations in halo EFT 
to $2n$ halo nuclei using a trimer auxiliary field formalism
and its application to the charge form factors and radii of
${}^{11}$Li, ${}^{14}$Be and ${}^{22}$C  \cite{Hagen:2013xga}.
Finally, we present results from a recent investigation of
Efimov physics in the Calcium isotope chain \cite{Hagen:2013jqa}.
This study combined $n$-${}^{60}$Ca S-wave scattering phase shifts 
from state of the art ab initio coupled cluster calculations using 
chiral interactions with halo EFT to obtain the properties of 
${}^{62}$Ca. In particular, correlations between 
different observables in the $n$-${}^{61}$Ca and ${}^{62}$Ca systems
were predicted and evidence of Efimov physics 
in ${}^{62}$Ca was provided. 

\setlength{\bibsep}{0.0em}
\begin{thebibliographynotitle}{99}

\bibitem{Bertulani-02}
  C.A.~Bertulani, H.-W.~Hammer and U.~Van Kolck,
  Nucl.\ Phys.\  A 712 (2002) 37
  [arXiv:nucl-th/0205063];
  P.F.~Bedaque, H.-W.~Hammer and U.~van Kolck,
  Phys.\ Lett.\  B 569 (2003) 159
  [arXiv:nucl-th/0304007].

\bibitem{Canham:2008jd}
  D.~L.~Canham and H.-W.~Hammer,
  Eur.\ Phys.\ J.\  A 37 (2008) 367
  [arXiv:0807.3258 [nucl-th]].

\bibitem{Acharya:2013aea} 
  B.~Acharya, C.~Ji and D.~R.~Phillips,
Phys.\ Lett.\ B 723 (2013) 196
  [arXiv:1303.6720 [nucl-th]].

\bibitem{Hagen:2013xga}
  P.~Hagen, H.-W.~Hammer and L.~Platter,
  arXiv:1304.6516 [nucl-th].

\bibitem{Hagen:2013jqa}
  G.~Hagen, P.~Hagen, H.-W.~Hammer and L.~Platter,
  arXiv:1306.3661 [nucl-th].

\end{thebibliographynotitle}

\newabstract 
\begin{center}
{\large\bf EFT for photon interactions with halo nuclei}\\[0.5cm]
{\bf Daniel Phillips} \\[0.3cm]
Institute for Nuclear and Particle Physics and Department of Physics and Astronomy,
Ohio University, Athens, OH 45701, USA\\[0.3cm]
\end{center}

``Halo EFT" contains core and neutron degrees of freedom, and is built
on the scale hierarchy $R_{\rm core} \ll R_{\rm
  halo}$~\cite{Hammer}. This theory can be used to analyze the
Coulomb dissociation 
of one-neutron halos. In ${}^{19}$C $R_{\rm core}/R_{\rm halo} \approx
0.4$, and fitting the N$^2$LO Halo EFT amplitude for the Coulomb
dissociation of ${}^{19}$C to experimental data allows accurate values
for the n$^{18}{\mathrm C}$ effective-range parameters   to be
extracted~\cite{AP13}. 

The ${}^{11}$Be nucleus has shallow $1/2^+$ and $1/2^-$ states. We
used data on the ${}^{11}$Be levels and the B(E1) of the $1/2^+$ to
$1/2^-$ transition to fix LO EFT parameters. We then predicted the
Coulomb dissociation spectrum of ${}^{11}$Be at LO. At next-to-leading
order an additional parameter associated with the asymptotic
normalization coefficient (ANC) of the $1/2^+$ state enters. It can be
adjusted to obtain a good description of data on the low-energy
dB(E1)/dE spectrum~\cite{HP11}. 

One may instead employ ANCs obtained in {\it ab initio} calculations
as input to Halo EFT. We have done this in the case of the reaction
${}^7{\rm Li} + n \rightarrow {}^8{\rm Li} + \gamma$~\cite{Zh13}. We
need 7 ANCs in order to fix the LO parameters of the theory that are
pertinent to radiative neutron capture into the ${}^8$Li ground and
first-excited states. Our LO result for the threshold ground-state
capture cross section is about 30\% below the data. The ratio of
ground-to-excited-state capture is within 1\% of the experimental
result, and we also obtain a good result for the ratio of capture from
different spin channels. This is in contrast to a previous EFT
calculation, which made simplifying assumptions about the reaction
dynamics~\cite{RH11}.  

p-wave 2n halos are also now being addressed in Halo EFT. A LO
calculation of ${}^6$He using the power counting of Ref.~\cite{Be03}
shows that a three-body force is necessary at LO in this
system~\cite{Ji13}. A similar conclusion, but using a different
approach and a different ${}^4$He-n amplitude, was reached in
Ref.~\cite{RvK12}. 

\setlength{\bibsep}{0.0em}
\begin{thebibliographynotitle}{99}  
  \bibitem{Hammer}
  H.-W. Hammer, these mini-proceedings.

\bibitem{AP13}
    B.~Acharya and D.~R.~Phillips,
Nucl.\ Phys.\ A {\bf 913}, 103 (2013). 

\bibitem{HP11} H.-W.~Hammer and D.~R.~Phillips,
  Nucl.\ Phys.\ A {\bf 865}, 17 (2011).
  
  \bibitem{NW11}
    K.~M.~Nollett and R.~B.~Wiringa,
  Phys.\ Rev.\ C {\bf 83}, 041001 (2011).
  
  \bibitem{Zh13}
  X.~Zhang, K.~Nollett, and D.~R.~Phillips, in preparation.
  
  \bibitem{RH11}
    G.~Rupak and R.~Higa,
  Phys.\ Rev.\ Lett.\  {\bf 106}, 222501 (2011).
  
  \bibitem{Be03}
    P.~F.~Bedaque {\it et al.},
  Phys.\ Lett.\ B {\bf 569}, 159 (2003).
  
  \bibitem{Ji13}
  C.~Ji, Ch.~Elster, and D.~R.~Phillips, in preparation.
  
  \bibitem{RvK12}
  J.~Rotureau and U.~van Kolck,
  Few Body Syst.\  {\bf 54}, 725 (2013).
\end{thebibliographynotitle}

\newabstract 
\begin{center}
{\large\bf Selected weak interaction processes on the deuteron and~$^3$He}\\[0.5cm]
{\bf Jacek Golak}, 
Alaa Eldeen Elmeshneb, 
Roman Skibi\'nski, \\
Kacper Topolnicki, 
and Henryk Wita{\l}a  \\[0.3cm]
M. Smoluchowski Institute of Physics, Jagiellonian University, \\
PL-30059 Krak\'ow, Poland\\[0.3cm]
\end{center}

In 1992 Walter Saake, a student of Prof. Walter Gl\'ockle, 
prepared his master thesis~\cite{saake}, in which he studied
numerical properties of the three-nucleon (3N) bound state and some
weak transitions between the $^3$H and $^3$He nuclei.
We now calculate selected decay rates for the muon-deuteron and 
muon-$^3$He atoms as well as the $ft$-value for the triton beta decay.
We use the single nucleon current operator without and with relativistic
corrections. We employ a new method to deal with partial wave decomposition (PWD)
of the current operator and show that for the two-nucleon (2N) system PWD can be totally
avoided~\cite{topolnicki}. We plan to include 2N current operators as given
for example in Refs.~[3-8].  
We do hope that the framework under construction
can be a useful tool for constructing a consistent treatment of forces and current operators 
within the chiral effective field theory.
 
\setlength{\bibsep}{0.0em}
\begin{thebibliographynotitle}{9}
\bibitem{saake} W. Saake, master thesis, Ruhr-Universit\"at, Bochum, (1992), unpublished.
\bibitem{topolnicki} K. Topolnicki {\it et al.}, Few-body Syst., DOI 10.1007/s00601-012-0479-y
\bibitem{marcucci1} L. E. Marcucci, Ph.D. thesis, Old Dominion
  University, (2002), \\
http://www.df.unipi.it/\~{}marcucci/
\bibitem{marcucci2} L. E. Marcucci {\it et al.}, Phys. Rev. {C}63 (2001) 015801.
\bibitem{ando} S. Ando et al., Phys. Lett. {B533} (2002) 25.
\bibitem{marcucci3} L. E. Marcucci {\it et al.}, Phys. Rev. {C}83 (2011) 014002.
\bibitem{marcucci4} L. E. Marcucci {\it et al.}, Phys. Rev. Lett. {108} (2012) 052502.
\bibitem{shen} G. Shen {\it et al.}, Phys. Rev. {C}86 (2012) 035503.
\end{thebibliographynotitle}

\newabstract 
\begin{center}
  {\large\bf High-accuracy analysis of Compton Scattering in $\chi$EFT}\\[0.5cm]
{\bf Harald W.~Grie{\ss}hammer}$^{1,2}$  \\[0.3cm]
$^1$ INS, 
The George Washington University, Washington, DC 20052, USA\footnote{Permanent
  address; e-mail: hgrie@gwu.edu}\\ $^2$ JCHP and IKP-3,
FZ J\"ulich, D-52428 J\"ulich, Germany\\[0.3cm]
\end{center}

Compton scattering from protons and neutrons provides important insight into
the structure of the nucleon. A new extraction of the static electric and
magnetic dipole polarisabilities ($\alpha_{E1}$ and $\beta_{M1}$) of the
proton and neutron from a new statistically consistent database taking into
account all published elastic data below 300~MeV in Chiral Effective Field
Theory finds (in units of $10^{-4}\,$fm$^3$):
\begin{eqnarray*}
  \mbox{proton~\cite{higherorderpols}: } \alpha_{E1}^{(p)}&=&10.7\pm0.4(\mathrm{stat})\pm0.2(\mathrm{Baldin})
  \pm0.3(\mathrm{theory})
  \\
  \beta_{M1}^{(p)} &=&
  3.1\mp0.4(\mathrm{stat})\pm0.2(\mathrm{Baldin})\mp0.3(\mathrm{theory})\\
  \mbox{neutron~\cite{comptonreview}: } \alpha_{E1}^{(n)}&=&11.1\pm
  1.8(\mathrm{stat})\pm0.4(\mathrm{Baldin})\pm0.8(\mathrm{theory})
  \\
  \beta_{M1}^{(n)} &=& 4.2\mp
  1.8(\mathrm{stat})\pm0.4(\mathrm{Baldin})\pm0.8(\mathrm{theory})
\end{eqnarray*}
Special care has been taken to reproducibly justify a theoretical uncertainty
of $\pm0.3$ from the most conservative of several estimates of higher-order
terms.  Within the statistics-dominated errors, the proton and neutron
polarisabilities are thus identical, i.e.~no isospin breaking effects of the
pion cloud are seen, as predicted by Chiral EFT.  An explicit $\Delta(1232)$
is particularly important for deuteron Compton scattering above about $90$~MeV
as measured at SAL and MAXlab. For few-nucleon systems like the deuteron and
${}^3$He, consistency arguments dictate that the $NN$ and $NNN$ rescattering
states must be included for a correct Thomson limit. In view of ongoing and
planned efforts at HI$\gamma$S, MAMI and MAXlab, single- and doubly-polarised
observables with linearly or circularly polarised photons on both un-, vector-
and tensor-polarised deuterons are important~\cite{dcomptonobs}. Several
observables can be used to extract not only scalar nucleon polarisabilities,
but also the so-far practically un-determined spin polarisabilities. These
parametrise the stiffness of the nucleon's low-energy spin degrees of freedom
in electro-magnetic fields, i.e.~the optical activity of the nucleon.

Supported  by US DoE DE-FG02-95ER-40907 and the Sino-German CRC 110.

 \vspace*{-0.3cm}

\setlength{\bibsep}{0.0em}
\begin{thebibliographynotitle}{99}
\bibitem{higherorderpols} J.~A.~McGovern, D.~R.~Phillips and
  H.~W.~Grie{\ss}hammer: 
  Europ.\ J.\ Phys.~\textbf{A}\textbf{49} (2013) 12 [arXiv:1210.4104 [nucl-th]].

\bibitem{comptonreview} H.~W.~Grie{\ss}hammer, J.~A.~McGovern, D.~R.~Phillips and
  G.~Feldman: 
  Prog.~Part.~Nucl.~Phys.~\textbf{67} (2012), 841 [arXiv:1203.6834 [nucl-th]].

\bibitem{dcomptonobs} H.~W.~Grie{\ss}hammer: 
  Europ.~J.~Phys.~\textbf{A} in press [arXiv:1304.6594 [nucl-th]].

\end{thebibliographynotitle}

\newabstract 
\begin{center}
{\large\bf Chiral expansion of the three-nucleon force}\\[0.5cm]
{\bf A.M. Gasparyan}, H. Krebs, E. Epelbaum  \\[0.3cm]
 Institut f\"ur Theoretische Physik II, Ruhr-Universit\"at Bochum,\\
44780 Bochum, Germany\\  
 Institute for Theoretical and Experimental Physics,\\
 117218, B.~Cheremushkinskaya 25, Moscow, Russia
\end{center}

A precise, quantitative description of the three-nucleon forces (3NFs)
is needed to understand nucleon-deuteron elastic and breakup scattering
(in particular the $A_y$-puzzle ) 
and to improve \emph{ab-initio}  nuclear structure calculations.
We describe the calculation of the long-range contributions to the 3NF
up to next-to-next-to-next-to-next-to-leading order (N$^4$LO) in the 
$\Delta$-less chiral EFT framework. The long-range topologies do not involve free
parameters except for the low-energy constants (LECs) which have 
been extracted from $\pi N$  scattering  using the same theoretical
framework. 
%
To study convergence of the chiral expansion,  
we worked out the most general operator structure of a local
isospin-invariant 3NF which involves 89 independent
operators. We proposed a set of 22
operators which can serve as a basis and give rise to all 89 structures in the
3NF upon making permutations of the nucleon labels.
Using this operator basis, we compared the strength of the
corresponding profile functions  
in configuration space for individual topologies at various orders in the chiral expansion. 
We observe a good
convergence for the longest-range $2\pi$-exchange topology which
clearly dominates the 3NF at distances of the order $r\gtrsim 2$ fm.
The intermediate-range $2\pi$-$1\pi$ exchange and ring diagrams
provide sizable corrections  at $r\sim 1$ fm and 
contribute to those $12$ profile functions which vanish for the $2\pi$
exchange. As expected, we found that N$^4$LO corrections to the
intermediate-range topologies are numerically large and in most cases
dominate over the nominally leading N$^3$LO terms. This can be traced
back to the role played by the $\Delta$(1232) isobar whose excitations
provide an important 3NF mechanism.
For the intermediate-range topologies, 
first effects of the $\Delta$ appear at N$^4$LO through resonance
saturation of the LECs $c_2$,  $c_3$ and $c_4$.  
The
importance of the $\Delta$ isobar is reflected in the 
large values of these LECs which are responsible for large N$^4$LO
corrections we observe. Our preliminary results in the $\Delta$-full
approach demonstrate indeed a better convergence pattern for the 3NF.
We also observe that some 3NF operators receive sizable contributions 
from double $\Delta$-excitation graphs. Such contributions are absent 
in the $\Delta$-less N$^4$LO potential, which may indicate that the
$\Delta$-full approach  is more efficient.

\setlength{\bibsep}{0.0em}
\begin{thebibliographynotitle}{99}
\bibitem{Krebs:2012yv} 
  H.~Krebs, A.~Gasparyan, E.~Epelbaum,
  Phys.\ Rev.\ C {\bf 85}, 054006 (2012).
\bibitem{Krebs:2013kha} 
  H.~Krebs, A.~Gasparyan E.~Epelbaum,
  Phys.\ Rev.\ C {\bf 87}, 054007 (2012).
 \end{thebibliographynotitle}

\newabstract 
\begin{center}
{\large\bf Three-nucleon forces in the $1/N_c$ expansion }\\[0.5cm]
{Daniel Phillips}$^1$, and {\bf Carlos Schat}$^{1,2}$  \\[0.3cm]
$^1$ Institute of Nuclear and Particle Physics and 
Department of Physics and Astronomy, Ohio University, Athens, Ohio 45701, USA;\\[0.3cm]
$^2$ CONICET - Departamento de F\'{\i}sica, FCEyN, Universidad de Buenos Aires, 
Ciudad Universitaria, Pab.~1, (1428) Buenos Aires, Argentina.
\end{center}

We present a complete classification of all the spin-flavor structures that can contribute
to three-nucleon forces and power count them in $1/N_c$. Including all independent momentum
structures,  a complete basis of operators for the three-nucleon force is given explicitly
up to next-to-leading order in the $1/N_c$ expansion. We also show that the expansion is
in a power series in $1/N_c^2$. This summarizes recent findings that have been presented
in Ref.~\cite{Phillips:2013rsa}.

In the context of nuclear forces the $1/N_c$ expansion was first used to study the central
part of the NN potential by Savage and Kaplan~\cite{KS96}, and then to analyze the
complete potential, classifying the relative strengths of the central, spin-orbit and
tensor forces, by Kaplan and Manohar~\cite{KM97}.  To obtain the results mentioned above
we extended these analyses of the two-nucleon force to the case of the three-nucleon
force (3NF). 

In particular, we find that at leading order in $1/N_c$ a spin-flavor independent term is
present, as are the spin-flavor structures associated with the Fujita-Miyazawa
three-nucleon force.  Modern phenomenological three-nucleon forces like the Urbana
potential \cite{Pi01} are thus consistent with this ${\cal O}(N_c)$ leading force,
corrections to which are suppressed by $1/N_c^2$.

We obtain another interesting result if we restrict our basis to a subset of time-reversal
even operators: we find a total of 80 operators that constitute the most general  basis
for a local 3NF.  In a recent paper~\cite{Krebs:2013kha} the authors needed a basis of 89
operators to obtain the most general contribution of a local 3NF.  An important subject
for future investigation is the relation between the two sets of operators, and a
determination of the minimal basis of operators for a general, local 3NF.

\setlength{\bibsep}{0.0em}
\begin{thebibliographynotitle}{99}

\bibitem{Phillips:2013rsa} 
  D.~R.~Phillips and C.~Schat,
  arXiv:1307.6274 [nucl-th]. 

\bibitem{KS96}
  D.~B.~Kaplan and M.~J.~Savage,
  Phys.\ Lett.\ B {\bf 365}, 244 (1996).

\bibitem{KM97}
  D.~B.~Kaplan and A.~V.~Manohar,
  Phys.\ Rev.\  C {\bf 56}, 76 (1997).

\bibitem{Pi01}
    S.~C.~Pieper, V.~R.~Pandharipande, R.~B.~Wiringa and J.~Carlson,
  Phys.\ Rev.\ C {\bf 64}, 014001 (2001).

\bibitem{Krebs:2013kha}
  H.~Krebs, A.~Gasparyan and E.~Epelbaum,
  Phys.\ Rev.\ C {\bf 87}, 054007 (2013).

\end{thebibliographynotitle}

\newabstract 
\begin{center}
{\large\bf Constraining the three-nucleon contact interaction from nucleon-deuteron elastic scattering}\\[0.5cm]
{\bf Luca Girlanda}$^1$, Alejandro Kievsky$^2$, and Michele Viviani$^2$  \\[0.3cm]
$^1$Dip. di Matematica e Fisica, Universit\`a del Salento,
and INFN Lecce, Italy\\[0.3cm]
$^2$INFN  Pisa, Italy\\[0.3cm]
\end{center}

None of the presently available models of the three-nucleon ($3N$)
interaction leads to a satisfactory description of bound and
scattering states of $A=3$ systems. It seems natural to ascribe the
above situation to the fact that these models include a very small
number of adjustable parameters, compared to the two-nucleon
interaction case. Moreover, the main discrepancies arise at very low
energy, as in the case of the proton vector analyzing power $A_y$ in
$p-d$ scattering. At such low energies, much smaller than $M_\pi$, the
interactions among the three nucleons are effectively of contact
nature. We therefore propose to include the subleading $3N$ contact
terms, which contain two powers of nucleon momenta and are
unconstrained by chiral symmetry. By using all constraints from
discrete and space-time symmetries we arrived at a set of 10
independent operators, and parametrized the potential in terms of 10
LECs, $E_{1,...,10}$ \cite{gkm}. The basis of operators has been
chosen such that most terms in the potential can be viewed as an
ordinary interaction of a pair of particles with a further dependence
on the coordinate of the third particle. In particular, the terms
proportional to $E_7$ and $E_8$ are of spin-orbit type, and, as
already suggested in the literature \cite{kievsky99}, have the right
properties to solve the $A_y$ puzzle. Limiting ourselves to contact
interactions, we thus have 11 LECs, a leading one $E=c_E/(F_\pi^4
\Lambda)$, and ten subleading ones $E_i=e^{3N}_i/(F_\pi^4 \Lambda^3)$,
with adimensional quantities $c_E, e^{3N}_i \sim O(1)$, according to
naive dimensional analysis. With the above $3N$ potential, used in
conjunction with the AV18 two-nucleon interaction, we obtain the $p-d$
phaseshifts using the HH method \cite{kievsky08}, and compare to the
phaseshift analysis (PSA) \cite{kievsky96}. We observe a strong
sensitivity to $c_E$, $e^{3N}_8$ and $e^{3N}_{10}$, and fit them (for
a given $\Lambda = 500$~MeV) to reproduce the P-waves $^4 P_{1/2}$,
$\epsilon^-_{3/2}$ and $^4 P_{5/2}$  (particularly important for the
$A_y$ problem), finding natural values of the LECs but a large
$\chi^2$. This may be  due to the optimistic experimental errorbars of
the energy-independent PSA \cite{kievsky96}. Work  is in progress to
optimize the fitting procedure and explore the sensitivity to
$\Lambda$.

\setlength{\bibsep}{0.0em}
\begin{thebibliographynotitle}{99}
\bibitem{gkm}  L.~Girlanda, A.~Kievsky and M.~Viviani,
  Phys.\ Rev.\ C 84 (2011) 014001.
\bibitem{kievsky99}   A.~Kievsky,
  Phys.\ Rev.\  C 60  (1999) 034001.
\bibitem{kievsky08}
A.~Kievsky et al., J. Phys. G 35 (2008) 063101.
\bibitem{kievsky96}
A.~Kievsky et al., Nucl. Phys. A 607 (1996) 402.
\end{thebibliographynotitle}

\newabstract 
\begin{center}
{\large\bf Exploring Three Nucleon Forces in Few Nucleon Scattering}\\[0.5cm]
{\bf Kimiko Sekiguchi}  \\[0.3cm]
Department of Physics, Tohoku University, Sendai, 980-8578, Japan
\\[0.3cm]
\end{center}

Experimentally, one must utilize systems with more than two nucleons
($A\ge3$) to explore the properties of three nucleon forces (3NFs).
Three nucleon scattering is one of the most promising tools,
because this system provides a rich set of energy dependent 
spin observables and cross sections.
In 1998 two theory groups incorporated 3NFs in elastic 
nucleon--deuteron ($Nd$) scattering 
at intermediate energies ($E \gtrsim 60$ MeV/nucleon),
and they suggested that the difference found in the cross section minima
is the signature of 3NF effects~\cite{wit98}, \cite{nemoto98}.
Since then we have extensively performed experimental studies of 
intermediate--energy $pd$ and $nd$ scattering at RIKEN
and RCNP~\cite{riken}, \cite{rcnp}, 
providing precise data for cross sections and a variety of spin observables.  
Recently we have extended the measurements
at the new facility of RIKEN RI beam factory~\cite{sek2011}
where polarized deuteron beams are available up to 400 MeV/nucleon.

The results of comparison between elastic scattering data
and the state--of--the--art Faddeev calculations based on 
realistic nucleon--nucleon forces plus $2\pi$--exchange 3NFs 
are summarized as follows;
(i)
A clear signature of 3NFs is identified 
in the cross section minimum for $Nd$ elastic scattering. 
(ii)
The polarization observables are not always
described by adding the 3NFs, 
indicating defects of spin dependent parts of 3NFs.
(iii)
As going to higher incident energies  ($\gtrsim 200$ MeV/nucleon)
the serious discrepancies between the data and the 
calculations appear at the backward angles,
which are not remedied even by including the 3NFs.
Some significant components are missing 
in the higher momentum transfer region.

As the next step of experimental study of 
few nucleon scattering it should be interesting to 
extend the measurements to 4N scattering systems, which
are the first step from few to many body systems,
and to obtain 3N scattering data at higher energies 
to investigate higher momentum components of nuclear forces.

\setlength{\bibsep}{0.0em}
\begin{thebibliographynotitle}{99}
\bibitem{wit98}
H. Wita{\l}a {\it et al.}, 
Phys. Rev. Lett. {\bf 81} (1998) 1183.
\bibitem{nemoto98}
S. Nemoto {\it et al.}, Phys.\ Rev.\ C {\bf 58}(1998) 2599.
\bibitem{riken}
For example, 
K.\ Sekiguchi {\it et al.}, Phys.\ Rev.\ C\ {\bf 65} (2002) 034003;
{\it ibid.} {\bf 79} (2009) 054008; Phys. Rev. Lett. {\bf 95} (2005) 162301.
\bibitem{rcnp}
K.\ Hatanaka {\it et al.}, Phys.\ Rev.\ C {\bf 66} (2002)  044002;
Y.\ Maeda {\it et al.}, {\it ibid.} {\bf 76} (2007)  014004.
\bibitem{sek2011}
K.\ Sekiguchi {\it et al.}, Phys.\ Rev.\ C {\bf 83} (2011)  061001.
\end{thebibliographynotitle}

\newabstract 
\begin{center}
{\large\bf Study of nuclear forces at intermediate energies}\\[0.5cm]
{\bf Nasser Kalantar-Nayestanaki}  \\[0.3cm]
KVI, University of Groningen,\\
Groningen, The Netherlands\\[0.3cm]
\end{center}

At KVI and many other laboratories, various combinations of high-precision cross sections, 
analyzing powers, spin-transfer and spin-correlation coefficients have been measured at different 
incident proton or deuteron beam energies between 100 and 200 MeV for a large range of 
scattering angles. These measurements have been performed for elastic and break-up 
reactions as well as for the radiative-capture process. Calculations 
based on two-body forces only do not describe the data sufficiently. The inclusion of 
three-body forces improves the discrepancies with the data significantly. However, there 
are still clear deficiencies in the calculations \cite{reviewarticle}.

The question that arises when looking at the bulk of the data and the remaining discrepancies 
which are sometimes sizable \cite{ramazani} is: what are the sources of the lack of 
understanding of the three-nucleon systems? Considering the fact that the calculations are 
numerically exact, the only remaining possibility is our understanding of the nuclear force 
itself. Although there is a lot of progress in producing consistent nuclear potentials within
the framework of effective field theories, one has to look deeper into this problem. In 
fact, one can argue that part of the problem in the nuclear potentials, presently on the market, is
that they all use experimental observables primarily from elastic nucleon-nucleon scattering. In
that way, the sensitivity to the off-shell effects of these potentials could be masked. These off-shell
properties should show up again in the calculations of observables in three-body systems. Precise
measurements of nucleon-nucleon bremsstrahlung process have also presented major disagreements
with the potential model calculations in the past
\cite{bremsstrahlung1}, \cite{bremsstrahlung2}. The only
complication in this process is the presence of the electromagnetic operator. Recent developments in
effective field theories indicate that this operator will be under control soon. Once this is the case,
it is mandatory to revisit the bremsstrahlung process and see whether one understands this process as 
well alongside the three-nucleon system.

\setlength{\bibsep}{0.0em}
\begin{thebibliographynotitle}{99}
\bibitem{reviewarticle} N. Kalantar-Nayestanaki, E. Epelbaum, J.G. Messchendorp and A. Nogga,
Rep. Prog. Phys. 75 (2012) 016301 and references therein. 
\bibitem{ramazani} A. Ramazani-Moghaddam-Arani et al., Phys. Rev. C78 (2008) 014006.

\bibitem{bremsstrahlung1} M. Mahjour-Shafiei et al., Phys. Rev. C70 (2004) 024004 and references therein.
\bibitem{bremsstrahlung2} M. Mahjour-Shafiei et al., Phys. Lett. B632 (2006) 480.

\end{thebibliographynotitle}

\newabstract 
\begin{center}
{\large\bf Light hypernuclei based on chiral interactions at next-to-leading order}\\[0.5cm]
{\bf Andreas Nogga}  \\[0.3cm]
Institute for Advanced Simulation (IAS-4), Institut f\"ur Kernphysik (IKP-3), and J\"ulich Center for Hadron Physics  (JCHP), 
Forschungszentrum J\"ulich,\\
D-52425 J\"ulich, Germany\\[0.3cm]
\end{center}

Chiral perturbation theory is an important tool to develop consistent two- and more-baryon 
interactions based on the symmetries of Quantum Chromo Dynamics. 
Here, we  discuss predictions for the binding energies of the light hypernuclei 
$^3_\Lambda$H, $^4_\Lambda$He and $^4_\Lambda$H based on chiral hyperon-nucleon interactions
\cite{Noggaprep:2013}. 
These calculations are based on leading and next-to-leading order chiral hyperon-nucleon (YN)
interactions \cite{NoggaPolinder:2006}, \cite{NoggaHaidenbauer:2013}.  We briefly introduce these 
interactions and their description of the YN low energy data. Then we use the Faddeev-Yakubovsky 
technique of Ref. \cite{NoggaNogga:2002} to obtain first predictions  
for light hypernuclei based chiral interactions at next-to-leading order. 
The charge-symmetry breaking of $^4_\Lambda$He and $^4_\Lambda$H is studied in some detail.

\setlength{\bibsep}{0.0em}
\begin{thebibliographynotitle}{99}
\bibitem{Noggaprep:2013} J. Haidenbauer,  U.-G. Mei{\ss}ner, and A. Nogga, in preparation. 
\bibitem{NoggaPolinder:2006} H. Polinder, J. Haidenbauer, and U.-G. Mei{\ss}ner, Nucl. Phys. {\bf A 779} (2006) 244.
\bibitem{NoggaHaidenbauer:2013} J. Haidenbauer, S. Petschauer,
  N. Kaiser, U.-G. Mei{\ss}ner, A.Nogga, W. Weise, Nucl. Phys. {\bf A
    915} (2013)  24.
\bibitem{NoggaNogga:2002}  A. Nogga, H. Kamada, and W. Gl\"ockle, Phys. Rev. Lett. {\bf 88}  (2002) 172501.
\end{thebibliographynotitle}

\newabstract 
\begin{center}
{\large\bf
The reaction $K^- d \to \pi \Sigma n$ in the $\Lambda$(1405) resonance region
}\\[0.5cm]
{\bf Kazuya Miyagawa}$^1$, and Johann Haidenbauer$^2$  \\[0.3cm]
$^1$Simulation Science Center, Okayama University of Science,\\
1-1 Ridai-cho, Okayama 700-0005, Japan\\[0.3cm]
$^2$Institute for Advanced Simulation,\\
Forschungszentrum J\"ulich, D-52425 J\"ulich, Germany\\[0.3cm]
\end{center}
The present work investigates the reaction $K^- d \to \pi \Sigma n$~\cite{R1}.
It is motivated by a corresponding proposal for an experiment at the 
J-PARC 50-GeV proton synchrotron whose primary goal is to study the mass and width of 
the $\Lambda$(1405) resonance 
 \cite{jparc}.
As in Ref.~\cite{R1}, 
we take into account single and two-step processes owing to $\bar K N \to \pi \Sigma$ rescattering
and evaluate the $\pi \Sigma$ invariant mass spectrum in the $\Lambda$(1405) resonance region.

A similar calculation was already performed by Jido et al.~\cite{Jido09}.
We scrutinize the results of Ref.~\cite{Jido09}, \cite{Jido12} avoiding
some of the approximations introduced in those articles.
In addition, we consider elementary $\bar KN$-$\pi \Sigma$ amplitudes
generated from different interaction models, namely besides the one of
the Oset-Ramos model~\cite{ORB02} employed in \cite{Jido09}, \cite{Jido12} 
also those of a coupled-channel meson-exchange interaction \cite{MG}, \cite{Hai11}
that likewise predicts two poles in the region of the $\Lambda(1405)$ resonance.
We find that the $\pi \Sigma$ invariant mass spectra are suppressed by 
the deuteron wave function and the fall-off of the $n K^- p$
Green's function,
and show no clear peaks below the $n K^- p$  threshold
for both potentials.

However, as is pointed out in Ref.~\cite{R1}, one cannot rule out
the possibility that peaks would be generated by the inclusion 
of all rescattering processes summed up to infinite order
and one should rather rely on Faddeev-type approaches.
Now we continue our study in that direction, and
preliminary results incorporating in addition $\bar K N$ rescattering
suggest that the truncation of the multiple scattering series
adopted so far could be inadequate.

\setlength{\bibsep}{0.0em}
\begin{thebibliographynotitle}{99}
\bibitem{R1} 
 K.~Miyagawa, J.~Haidenbauer,
 Phys.\ Rev.\ C {\bf 85} (2012) 065201; \\
 Few-Body Syst. 2013, DOI 10.1007/s00601-013-0657-6.
\bibitem{jparc}
 S. Ajimura et al., {\rm http://j-parc.jp/NuclPart/pac\_0907/pdf/Noumi.pdf}
\bibitem{Jido09}
        D. Jido, E. Oset, T. Sekihara,
        Eur. Phys. J. A {\bf 42} (2009) 257.
\bibitem{Jido12}
  D.~Jido, E.~Oset, T.~Sekihara,
  arXiv:1207.5350 [nucl-th].
\bibitem{ORB02}
        E. Oset, A. Ramos, C. Bennhold, Phys. Lett. B {\bf 527} (2002) 99.
\bibitem{MG}
        A.~M\"uller-Groeling, K. Holinde, J. Speth,
        Nucl. Phys. A {\bf 513} (1990) 557.
\bibitem{Hai11}
  J.~Haidenbauer, G.~Krein, U.-G.~Mei{\ss}ner, L.~Tolos,
  Eur.\ Phys.\ J.\ A {\bf 47} (2011) 18.
\end{thebibliographynotitle}

\newabstract 
\begin{center}
{\large\bf The chiral electromagnetic currents applied to the deuteron and $^3$He disintegrations}\\[0.5cm]
{\bf Roman Skibi\'nski} \\[0.3cm]
M.Smoluchowski Institute of Physics, Jagiellonian University, PL-30059 \\
Krak\'ow, Poland,\\[0.3cm]
\end{center}

The Chiral Effective Field Theory delivers a systematic way to describe 
the nucleon-nucleon as well as the three nucleon interactions \cite{2}.
This formalism has been extended to electromagnetic interactions and the 
nuclear electromagnetic current has been derived in \cite{3}, \cite{4}. Up to NLO many 
different topologies contribute to the electromagnetic current what results 
in many momentum dependent spin-isospin operator structures. Inclusion of 
electromagnetic currents into few-body calculations in a scheme of
\cite{5}, \cite{6} 
requires their partial wave decomposition. The efficient, semi-automatized 
method of such decomposition was proposed in \cite{7} for two- and three-body 
operators and applied to the electromagnetic currents in \cite{8}.  

In Ref \cite{8} the deuteron and $^3$He photodisintegrations were studied with 
different parametrizations of the chiral nuclear potential at N$^2$LO \cite{2}. 
The electromagnetic current was constructed from the single-nucleon current, 
the one-pion exchange current and the leading two-pion exchange currents at NLO. 
The resulting predictions for the deuteron photodisintegration show, in general, 
good agreement with the data and with the predictions based on AV18 interaction
at photon energies below 70 MeV. 
While at higher energies and for the differential cross section the band which 
origins in different parametrizations of chiral forces becomes broad, 
for the polarized observables this band remains reasonably narrow.

For the $^3$He photodisintegration the width of bands grows with the increasing 
photon energy what do not allow for quantitative conclusions above E$_\gamma$=20 MeV. 
This points that the short-range meson exchange currents at NLO have to 
be included in the future theoretical analysis. 

\setlength{\bibsep}{0.0em}
\begin{thebibliographynotitle}{99}

\bibitem{2} E.Epelbaum, Prog. Part. Nucl. Phys. 57 (2006) 654.
\bibitem{3} S.K\"oling {\it et al.}, Phys. Rev. C80 (2009) 045502.
\bibitem{4} S.K\"oling {\it et al.}, Phys. Rev. C84 (2011) 054008. 
\bibitem{5} J.Golak {\it et al.}, Phys. Rept. 415 (2005) 89.
\bibitem{6} R.Skibi\'nski {\it et al.}, Eur. Phys. J. A24 (2005) 31.
\bibitem{7} J.Golak {\it et al.}, Eur. Phys. J. A43 (2010) 241.
\bibitem{8} D.Rozp\c edzik {\it et al.}, Phys. Rev. C83 (2011) 064004.
\end{thebibliographynotitle}

\newabstract 
\begin{center}
{\large\bf Muon capture  connecting to the three-nucleon force}\\[0.5cm]
U. Raha\footnote{Permanent address: 
Dept. Physics, I.I.T., Guwahati, 781 039 Assam, India}, 
S. Pastore, {\bf F. Myhrer}, and K. Kubodera  \\[0.3cm]
Dept. Physics and Astronomy, University of South Carolina, \\
Columbia, S.C. 29208, USA\\[0.3cm]
\end{center}

The rate of muon capture in a muonic hydrogen atom
is calculated in heavy-nucleon chiral perturbation theory 
(HB$\chi$PT) 
up to next-to-next-to leading order   
where
corrections due to the QED 
and the proton-size effect are included. 
Since the low-energy constants (LECs) involved
are determined from other independent sources of information,
the theory has predictive power.
For the hyperfine-singlet $\mu p$ capture rate $\Gamma$,
our calculations~\cite{ando2000}, \cite{udit2012}, \cite{saori2013}  give
$\Gamma=713 \,\pm 4\pm 2\pm 1\,s^{-1}$, 
where the uncertainties are due to the (historical) 
variations in the values of
$g_A$, $g_{\pi NN}$ and the proton axial radius 
$\langle r_A^2 \rangle$, respectively. 
The estimated next order HB$\chi$PT correction is about 1\%. 
The value quoted is in excellent agreement with 
the experimental value~\cite{MuCap2013}. 

The MuSun Collaboration~\cite{MuSun} is measuring the muon capture on  deuteron.
Given the accuracy of HB$\chi$PT, Ref.~\cite{MuSun} aims at 
determine an unknown  two-nucleon-pion  LEC   
which also enters  the chiral three-nucleon  potential  
 as well as the 
reaction $pp \to NN\pi$~\cite{vadim}.
Furthermore, this LEC also affects, e.g.,    
the primary $pp$ solar fusion reaction and $\nu d$ 
reactions which were used to determine the 
solar neutrino flux at the Sudbury Neutrino Observatory.  
See the muon capture review~\cite{kammelkubodera}  and recent theoretical work~\cite{mudeut}.

\setlength{\bibsep}{0.0em}
\begin{thebibliographynotitle}{99}
\bibitem{ando2000}
S. Ando, F. Myhrer and K. Kubodera, Phys. Rev. C {\bf 63} (2000) 015203.
\bibitem{udit2012}
U. Raha, F. Myhrer and K. Kubodera, Phys. Rev. C {\bf 87} (2013) 055501. 
\bibitem{saori2013}
S. Pastore, F. Myhrer and K. Kubodera, in prepraration.
\bibitem{MuCap2013}
V.A. Andreev {\it et al.} 
Phys. Rev. Lett. {\bf 110} (2013) 012504.
 
\bibitem{MuSun} V.A. Andreev {\it et al.} arXiv:1004.1754[nucl-ex]. 
\bibitem{vadim} V. Baru, invited talk at this workshop. 
\bibitem{kammelkubodera}
P. Kammel and K. Kubodera, Annu.Rev.Nucl.Part.Sci. {\bf 60} (2010) 327. 
\bibitem{mudeut} L.E. Marcucci {\it et al.} Phys. Rev. Lett. {\bf108} (2012) 052502;
Adam {\it et al.} Phys. Lett. B {\bf 709} (2012) 93; Y.-H. Song {\it et al.} to appear. 
\end{thebibliographynotitle}

\newabstract 
\begin{center}
{\large\bf Towards (d,p) reactions with Heavy Nuclei in a Faddeev Description}\\[0.5cm]
{\bf Ch. Elster}$^1$, L. Hlophe$^1$, N. Upadhyay$^2$, V. Eremenko$^1,6$, R.C. Johnson$^3$, 
F.M. Nunes$^1$,
G. Arbanas$^4$, J.E. Escher$^5$, I.J. Thompson$^5$,  \\[0.3cm]
$^1$Institute of Nuclear and Particle Physics,
 Ohio University Athens, OH,  USA \\[0.1cm]
$^2$National Superconducting Cyclotron Laboratory, 
MSU, East Lansing, MI, USA \\[0.1cm]
$^3$Department of Physics, University of Surrey, Guildford, GU2 7XH, UK \\[0.1cm]
$^4$Nuclear Science and Technology Division, ORNL, Oak Ridge, TN
 USA \\[0.1cm]
$^5$Lawrence Livermore National Laboratory L-414, Livermore, CA,  USA \\[0.1cm]
$^6$D.V. Skobeltsyn Inst. of Nucl. Phys, 
Moscow State University, Moscow, RU \\[0.3cm]

\end{center}

Current interest in nuclear reactions, in particular with rare
isotopes, concentrates  on their reactions with neutrons. In order to facilitate
the study those reactions, indirect methods using deuterons, like (d,p) reactions must
be used. Those may be viewed as three-body reactions and described with
Faddeev techniques. Here the three-body Hamiltonian governing the dynamics contains the
nucleon-nucleon force describing the deuteron, and phenomenological
neutron- and proton-nucleus optical potentials, which in turn are fit to a large body of
elastic scattering data force between the neutron (proton) and the nucleus.

The application of momentum space Faddeev techniques to nuclear reactions has been pioneered
in Ref.~\cite{Deltuva:2009fp}, and successfully applied to (d,p) reactions
for light nuclei. However, when extending these
calculations to heavier nuclei~\cite{Nunes:2011cv}, it becomes
 apparent that
techniques employed for incorporating the Coulomb interaction in Faddeev-type
calculations of reactions with light nuclei can not readily
be extended to the heaviest nuclei. Therefore, a new formulation for treating
(d,p) reactions with the exact inclusion of the Coulomb force
 as well as target excitation was formulated in Ref.~\cite{Mukhamedzhanov:2012qv}.
This new approach relies on a separable representation of the interparticle forces.

We present a separable representation of complex optical potentials suited to be employed
in momentum space Faddeev type calculations of (d,p) reactions~\cite{linda}. 
They are a generalization of the 
Ernst-Shakin-Thaler scheme in such a way that they
fulfill the reciprocity theorem.  

\vspace{-5mm}
\setlength{\bibsep}{0.0em}
\begin{thebibliographynotitle}{99}
\vspace{-3mm}
\bibitem{Deltuva:2009fp} A. Deltuva, A. Fonseca, Phys. Rev. C{\bf 79}, 0144606 (2009).

\bibitem{Nunes:2011cv} F. Nunes, A. Deltuva, Phys. Rev. C{\bf 84}, 034607 (2011); N. Upadhyay,
A. Deltuve, F. Nunes, Phys. Rev. C{\bf 85}, 054621 (2012).

\bibitem{Mukhamedzhanov:2012qv} A. Mukhamedzhanov, V. Eremenko, A. Sattarov, Phys. Rev. C{\bf
86}, 034001 (2012).

\bibitem{linda} L. Hlophe, Ch. Elster, R.C. Johnson, N. Upadhyay, F.M. Nunes, G. Arbanas, V.
Ermenko, J.E. Escher, I.J. Thompson, in preparation.
\end{thebibliographynotitle}

\newabstract 
\begin{center}
{\large\bf Experimental Studies of Deuteron-Proton Breakup\\[0.3cm]
        at Medium Energies}\\[0.5cm]
{\bf Stanis{\l}aw Kistryn}$^1$ and El\.zbieta Stephan$^2$  \\[0.3cm]
$^1$Institute of Physics, Jagiellonian University, 30-059 Krak\'ow, 
Poland\\[0.1cm]
$^2$Institute of Physics, University of Silesia, 40-007 Katowice,
Poland\\[0.3cm]
\end{center}

Exact modeling of all details of the few-nucleon system dynamics is
a crucial prerequisite for reaching full understanding of the nuclear 
interactions. Thorough studies of three-nucleon (3N) systems have led to 
a conclusion that a proper description of the experimental data cannot 
be achieved with the use of nucleon-nucleon (NN) forces alone. This 
indicates a necessity of including additional dynamics: subtle effects 
of suppressed degrees of freedom, introduced by means of genuine 
three-nucleon forces (3NF), or, for a long-time neglected, Coulomb force. 

Those findings would not be possible without a strong improvement of 
the experimental methods. New generation experiments in the middle-energy 
region employing high-resolution, multi-detector arrangements, provide 
data of unprecedented accuracy - see e.g.\  reviews~\cite{Sag10} 
and~\cite{Kal12}.
The ways of exploiting all advantages of such precision measurements are
demonstrated by discussing a sample experiment of the $^1$H($\vec{\rm d}$,pp)n
breakup reaction at 130 MeV~\cite{Kis13}. The project, carried out at KVI, 
Groningen, The Netherlands and FZ J\"ulich, Germany, provided very abundant
data set - around 4500 points for cross section and around 800 data points for
each of the five analyzing powers (vector A$_{\rm x}$, A$_{\rm y}$ and tensor
A$_{\rm xx}$, A$_{\rm yy}$, A$_{\rm xy}$,).

Confronting the experimental data with the theoretical predictions shows the
sensitivity of the cross sections to 3NFs and to Coulomb force effects, while
there is no sensitivity of the deuteron vector analyzing powers to any 
additional dynamics beyond the NN forces. The behavior of the tensor analyzing 
powers is rather complicated, showing discrepancies between the calculations 
and the experimental data which must be considered as indications of 
deficiencies in the spin part of the assumed models of the 3NFs.
Studies of the discrepancies as a function of different kinematical 
variables, extended over a range of several beam energies, might provide
signposts for improvements in the modeling of 3N system dynamics. 

\setlength{\bibsep}{0.0em}
\begin{thebibliographynotitle}{99}
\bibitem{Sag10} K.~Sagara, Few-Body Syst. 48 (2010) 59.
\bibitem{Kal12} N.~Kalantar-Nayestanaki {\it et al.}, Rep. Prog. Phys. 
75 (2012) 016301.
\bibitem{Kis13} St.~Kistryn, E.~Stephan, J. Phys. G: Nucl. Part. 
Phys. 40 (2013) 063101. 
\end{thebibliographynotitle}

\newabstract 
\begin{center}
{\large\bf Nd scattering calculation with Low-momentum potential
}\\[0.5cm]
{\bf H. Kamada}$^1$, H. Wita\l a$^2$, R. Okamoto$^1$, K. Suzuki$^1$ and M. Yamaguchi$^3$  \\[0.3cm]
$^1$Department of Physics, Faculty of Engineering, Kyushu Institute of Technology,\\
1-1 Sensuicho, Tobata, 804-8550 Kitakyushu, Japan\\[0.3cm]
$^2$M. Smoluchowski Institute of Physics, Jagiellonian University,\\
PL-30059 Krak\'ow, Poland \\[0.3cm]
$^3$Research Center for Nuclear Physics, Osaka University, \\
Ibaraki 567-0047, Japan\\[0.3cm]
\end{center}

The low-momentum nucleon-nucleon potentials (LMNN) have been developed\cite{Bogner} 
by a renormalization unitary transformation  from realistic potentials.
The LMNN potential has a parameter $\Lambda$ which restricts upper limit of momentum space.
Using the LMNN potential 
the binding energies of triton and alpha particle were calculated
\cite{Fujii} to investigate the dependence of $\Lambda$ in few-nucleon system. 
The study showed  that  a choice $\Lambda \ge $~5 fm$^{-1}$ 
is necessary to obtain the original binding energies.

Although in the case of the similarity renormalization group method, recent studies 
\cite{Jurgenson} prove that, inclusion of an induced three-body force 
arising from effects of many-body 
complex and the truncated 2-body space, brings them back to the original binding 
energies even under the choice $\Lambda \le $~2.3~fm$^{-1}$.
This is a good situation for the chiral effective field theory ($\chi$EFT) 
whose cut-off parameter has been taking 
the choice 2.3~$\le \Lambda_\chi \le $~3.0~fm$^{-1}$.
Because the three-body force of $\chi$EFT is consistently built up with the two-body
 force, the three-body force already contains the above-mentioned induced part.  
 
We calculated the Nd scattering observables at $E$=10 and 135 MeV using the LMNN potential. The similar paper is already published  ~\cite{Deltuva}, however, we recalculated them by using our LMNN potential. 
Even if $\Lambda$ was set to 3.0~fm$^{-1}$ and the induced three-body force was not included, only a small change was seen in almost all the scattering observables.

\setlength{\bibsep}{0.0em}
\begin{thebibliographynotitle}{99}
\bibitem{Bogner} 
S. K. Bogner {\it et al.}, Phys. Rep. {\bf 386}  (2003) 1.
\bibitem{Fujii} S. Fujii, E. Epelbaum, H. Kamada, R. Okamoto, K. Suzuki and W. Gl\"ockle, Phys. Rev. C{\bf 70}(2004) 024003; 
A. Nogga, S. K. Bogner and A. Schwenk, Phys. Rev. C{\bf 70} (2004) 061002(R).
\bibitem{Jurgenson} E. D. Jurgenson {\it et al.}, Phys. Rev. C {\bf 87} (2013) 054312; 
K. A. Wendt, Phys. Rev. C {\bf 87} (2013) 061001(R).
\bibitem{Deltuva} A. Deltuva, A. C. Fonseca, S. K. Bogner, Phys. Rev. C {\bf 77}
(2008)  024002.
\end{thebibliographynotitle}

\end{document}